\documentclass[aps,prx,floatfix,nofootinbib,onecolumn,superscriptaddress,longbibliography]{revtex4-2}
\usepackage{dsfont}

\usepackage{braket}
\usepackage{siunitx}

\usepackage{float}
\usepackage{amsthm}
\usepackage{amsmath}
\usepackage{amsfonts}
\usepackage{physics}
\usepackage{amssymb}
\usepackage{graphicx}
\usepackage{color}
\usepackage{accents}
\usepackage{bbold}
\usepackage{mathtools}
\usepackage[colorlinks=true, citecolor=blue]{hyperref}
\usepackage[normalem]{ulem}
\usepackage[shortlabels]{enumitem}

\makeatletter 
\newsavebox{\@brx}
\newcommand{\llangle}[1][]{\savebox{\@brx}{\(\m@th{#1\langle}\)}%
  \mathopen{\copy\@brx\kern-0.5\wd\@brx\usebox{\@brx}}}
\newcommand{\rrangle}[1][]{\savebox{\@brx}{\(\m@th{#1\rangle}\)}%
  \mathclose{\copy\@brx\kern-0.5\wd\@brx\usebox{\@brx}}}
\makeatother

\newtheorem{theorem}{Theorem}[section]

\newtheorem{proposition}[theorem]{Proposition}
\newtheorem{corollary}[theorem]{Corollary}

\theoremstyle{definition}

\theoremstyle{remark}
\newtheorem*{remark}{Remark}

\usepackage{listings} 

\usepackage{xcolor}

\definecolor{codegreen}{rgb}{0,0.6,0}
\definecolor{codegray}{rgb}{0.5,0.5,0.5}
\definecolor{codepurple}{rgb}{0.58,0,0.82}
\definecolor{backcolour}{rgb}{0.95,0.95,0.92}
\lstdefinestyle{mystyle}{
  backgroundcolor=\color{backcolour},   commentstyle=\color{codegreen},
  keywordstyle=\color{magenta},
  numberstyle=\tiny\color{codegray},
  stringstyle=\color{codepurple},
  basicstyle=\ttfamily\scriptsize,
  breakatwhitespace=false,         
  breaklines=true,                 
  captionpos=b,                    
  keepspaces=true,                 
  showspaces=false,                
  showstringspaces=false,
  showtabs=false,
  xleftmargin=0.02\textwidth,
  rulecolor=\color[RGB]{200,200,200},
  frame=bt,
  framextopmargin=2pt,
  framexbottommargin=2pt,
  framexleftmargin=10pt,
  tabsize=2
}
\begin{document}
\title{A constructive solution to the $n$-representability problem}
\author{Nicholas C. Rubin}
\email{nickrubin@google.com}
\affiliation{
             Google Quantum AI, 
             San Francisco, CA, 94105, USA}

\author{A. Eugene DePrince III}
\affiliation{
             Department of Chemistry and Biochemistry,
             Florida State University,
             Tallahassee, FL 32306-4390, USA}
\affiliation{
             Google Quantum AI, 
             San Francisco, CA, 94105, USA}
\begin{abstract}
We present an instantiation of the constructive solution to the fermionic quantum marginal problem by building the complete family of two-body witnesses at each level of a weighted sum-of-squares hierarchy. Our construction combines positive-semidefinite Gram matrices with explicit linear nullspace constraints that exactly cancel all higher-body contributions following a similar idea to the existing solution but made explicit in open source software.
We demonstrate the weighted sum-of-squares construction numerically by solving the associated semidefinite program and recovering the exact ground-state energy of a Richardson Hamiltonian in the three-fermion sector. We further prove that higher-degree particle-number multipliers are unnecessary when the full sum-of-squares generating algebra is used.
\end{abstract}
\maketitle

\tableofcontents

\section{Introduction}
The $n$-representability problem aims to describe the space of fermionic two-particle reduced density matrices ($2$-RDM) that can be obtained from an ensemble state of $n$ fermions. This space is usually characterized by enumerating witnesses that verify membership in the valid set. 
The importance of the problem has been recognized since the 1960s~\cite{RevModPhys.35.668} due to the fact that the $2$-RDM provides a more compact representation of a quantum system than the full wavefunction and it contains all the necessary information for computation of a molecule's energy and most equilibrium properties relevant to chemistry. In the quantum chemistry and math communities, progress was made formalizing partial solutions~\cite{garrod1964reduction, coleman1978reduced, erdahl1978representability, nakata2001variational, zhao2004reduced, nakata2008variational, PhysRevA.74.032501} and has since been largely organized around finding computationally efficient necessary conditions on the $2$-RDM to ensure the marginal can be obtained from a valid $n$-particle state. In 1995 the United States National Research Council deemed $n$-representability as one of the ten most prominent research challenges in quantum chemistry~\cite{national1995mathematical}. 

In quantum information the $n$-representability problem generalization from fermions to a broader set of quantum particles has played an important role in defining correlations and the structure of mixed states with respect to the spectrum of their marginals~\cite{klyachko2006quantum}. The quantum information community was able to provide a solution to the $1$-RDM pure-state variant of the $n$-representability problem~\cite{altunbulak2008pauli} and various low-dimensional pure-state quantum marginal solutions~\cite{bravyi2003requirements,higuchi2003one, PhysRevX.11.021039} along with the identification of the $n$-representability problem as QMA-complete~\cite{PhysRevLett.98.110503}. Furthermore, proofs of a convergent approximation hierarchy for certifying quantum correlations (NPA hierarchy)~\cite{navascues2008convergent} along with the connection to non-commutative polynomial optimization~\cite{pironio2010convergent} fully provided the structure for an ensemble $n$-representability solution. 

In this work we demonstrate how to use the infrastructure of Ref.~\cite{pironio2010convergent} to generate a complete solution of the $n$-representability problem. We rigorously show that each level of the weighted sum-of-squares NPA hierarchy~\cite{navascues2008convergent} can be used to generate all possible non-negative two-body witnesses and provide a simple linear algebra protocol, and software, to implement the solution. As discussed in Ref.~\cite{pironio2010convergent} the hierarchy is finite, and we prove termination at the level equal to the number of particles in the fermionic system. Our construction establishes completeness in two stages: 1) we provide linear nullspace equations that capture
every at-most-two-body witness admitting a degree-$d$ weighted
SOS representation, including joint cancellations between the
SOS and particle-number multiplier terms, 2) we prove that
at $d=n$ this weighted SOS cone contains every Hermitian,
particle-conserving, at-most-two-body operator that is
nonnegative on the $n$-particle subspace.
We demonstrate the weighted SOS strategy recovers canonical v2RDM constraints like $\mathrm{DQG}$ conditions and $\mathrm{DQGT_{1}T_{2}T_{2}'}$~\cite{nakata2008variational} conditions as restricted, non-general, subsets.

Our proofs reinforce the solution to the $n$-representability problem put forth independently by Mazziotti in 2012~\cite{PhysRevLett.108.263002}, 2023~\cite{PhysRevLett.130.153001}, and 2026~\cite{yt4s-dlwr} and demonstrate the key ingredients of those works can be cast in terms of the weighted sum-of-squares positivity ansatze. Specifically, we demonstrate that: 1) all witnesses are generated by a weighted sum of squares using the nullspace condition, described by Mazziotti~\cite{PhysRevLett.130.153001}, that is needed to instantiate the dual formulation~\cite{erdahl1979two, cances2006electronic}; and 2) the additional variance term of Ref.~\cite{yt4s-dlwr} is precisely the particle-number multiplier term in weighted sum-of-squares. Furthermore, we show how the nullspace condition of Mazziotti~\cite{PhysRevLett.130.153001} impacts the space of feasible $2$-RDMs in the primal problem. Using weighted SOS to tighten the existing $n$-representability solution has also been demonstrated in recent work from Nakata~\cite{nakata2026spurious} where they use the weighted sum-of-squares representation to prove that if a $2$-RDM satisfies the $(2,q)$-conditions it also satisfies the $q$-positivity conditions. While the existing solution is complete, and recently proved sufficient, here we cast the general solution in a symbolic form amenable to computation to aid in the designing of efficient computational methods for lower bound certificates in chemistry and condensed matter simulation. While we suspect components of the $n$-representability solution generated from wSOS ansatze are anticipated by Ref.~\cite{PhysRevLett.108.263002, PhysRevLett.130.153001, yt4s-dlwr, PhysRevA.85.062507}, we find it difficult to fully understand their solution. Thus we feel presenting a self-contained solution relying on the operator algebra approach of wSOS adds substantial expository benefit.

As a computational verification we use a Richardson-type pairing Hamiltonian in the
We focus on this model because the approximate three-positivity does not improve the lower bound certificate substantially beyond 2-positivity, highlighting the gap of traditional v2RDM constraints. This result is expected, but we use approximate $3$-positivity to highlight the
complete degree-three weighted SOS formulation recovers the
exact energy within solver accuracy. The results also afford the opportunity to derive how contraction relations between RDMs emerge from the multiplier polynomials in wSOS. Furthermore, we prove
that degree-at-most-two particle-number multipliers suffice
to preserve the supremal energy bound at every level $d \geq 2$
of the complete fermionic SOS hierarchy. Higher-degree
multiplier contributions can be absorbed into the SOS part
and degree-at-most-two multipliers, with an arbitrarily small
decrease in the lower-bound energy certificate.

Attendant to the theory, we provide the software we used to generate the degree-$d$ weighted SOS 2-particle witness as a component of the $p^{\dagger}q$ fermionic computer algebra library~\cite{liebenthal2025automated}. With minor modifications the code can be extended to arbitrary weighted SOS. The nullspace construction is not unique to fermions and applicable to other outer approximation methods. 

\section{SOS Witnesses and $n$-representability}
Prior to discussing the weighted SOS ansatz we first describe the protocol for constructing all 2-particle witnesses from a degree-$d$ SOS. After providing a proof of completeness we sketch the numerical protocol for generating a symbolic representation of all the witnesses. 
A normalized candidate set of low-order moments is
$n$-representable if and only if it is nonnegative with respect to every
Hermitian, particle-conserving, at-most-two-body operator
that is positive semidefinite on $\mathcal{H}_n$~\cite{RevModPhys.35.668, garrod1964reduction, PhysRevA.75.032502}.
For example, degree-6 operators that can be reconstructed exactly from $2$-RDM information due to cancellation conditions of degree-6 terms by anticommutation relations constitute just one set of necessary conditions on the 2-RDM~\cite{erdahl1978representability, PhysRevA.74.032501, zhao2004reduced, nakata2008variational}. 
Our goal is to construct all two-body witnesses admitting a
degree-$d$ SOS representation. Throughout, ``degree-$d$ SOS''
refers to generators of ladder degree at most $d$, whose
products have ladder degree at most $2d$. The nullspace
conditions cancel all terms of ladder degree greater than four.
We will show that the DQG, T1, T2', and (2,4) conditions can be derived within this framework.  

\subsection{Constructing SOS two-body witnesses}

To construct a 2-body witness we must parameterize a subset of the SOS operators that cancel terms of ladder degree greater than four. First,
we define the operators used to form the sums of squares. Let
$\mathfrak U_R$ denote the complex linear span of all finite
products of $a_1,\ldots,a_R,a_1^\dagger,\ldots,a_R^\dagger$,
including the identity. These operators act on the $R$-mode
fermionic Fock space and satisfy the canonical anticommutation
relations (CAR). Let $o_1,\ldots,o_L$ be the distinct normal-ordered monomials of
degree at most $d$, including $\mathbb 1$, with a fixed ordering of
the orbital indices. We define
\begin{align}
\mathcal V_d
  = \operatorname{Span}_{\mathbb C}\{o_1,\ldots,o_L\}
  \subseteq \mathfrak U_R
\end{align}
and collect the monomials into the vector
\begin{align}
\vec{o}_d =
\begin{pmatrix}
o_1\\
o_2\\
\vdots\\
o_L
\end{pmatrix}.
\end{align}
Every $R_k\in\mathcal V_d$ can therefore be written as
\begin{align}
R_k= \vec{c}_k  \cdot \vec{o}_d
   =\sum_{i=1}^L c_{ki}o_i,
\qquad
\vec{c}_k\in\mathbb C^{1\times L}.
\end{align}

The degree-$d$ SOS cone is
\begin{align}
\Sigma_d
 =\left\{
   \sum_{k=1}^r R_k^\dagger R_k
   \;:\;
   r<\infty,\ R_k\in\mathcal V_d
  \right\}.
\end{align}
Every element of $\Sigma_d$ is a globally positive-semidefinite
Hermitian operator on Fock space~\cite{helton2002positive}.
This definition has a Gram-matrix form. To see this, let
$W=\sum_{k=1}^r R_k^\dagger R_k$. Substitution of
$R_k= \vec{c}_k   \cdot \vec{o}_d$ gives
\begin{align}
W
 &= \vec{o}_d^{\dagger}
   \left(\sum_{k=1}^r \vec{c}_k^\dagger \vec{c}_k\right)
   \vec{o}_d \nonumber\\
 &= \vec{o}_d^{\dagger} G \vec{o}_d,
\qquad
G=\sum_{k=1}^r \vec{c}_k^\dagger \vec{c}_k\succeq0.
\end{align}
Conversely, any $G\succeq0$ can be factored as $G=C^\dagger C$.
The rows of $C$ define operators $R_k\in\mathcal V_d$, so every
positive semidefinite Gram matrix also produces an element of
$\Sigma_d$. Thus, writing
\begin{align}
\Phi_d(G):=\vec{o}_d^{\dagger} G \vec{o}_d,
\end{align}
we have the equivalent parameterization of $\Sigma_{d}$
\begin{align}
\Sigma_d=\{\Phi_d(G):G\succeq0\}.
\end{align}

To determine the subset of $\Sigma_{d}$ that can be evaluated with just $2$-RDM information we need to specify the structure of $G$ which satisfies the constraint. Because each $o_i$ has degree at most $d$, the product
$o_i^\dagger o_j$ can contain terms of degree up to $2d$ after
normal ordering. 
The canonical
anticommutation relations give an expansion of each product in
a normal-ordered basis:
\begin{align}
o_i^\dagger o_j
 =\sum_{m=0}^{2d}\sum_{\gamma\in\mathcal I_m}
   \Gamma_{ij}^{(m,\gamma)}O_\gamma^{(m)},
\end{align}
where $\mathcal I_m$ labels the distinct normal-ordered operators
$O_\gamma^{(m)}$ of degree $m$. The coefficients
$\Gamma_{ij}^{(m,\gamma)}$ are determined by normal ordering
$o_i^\dagger o_j$ under the application of Wick's theorem with respect to the true vacuum $|0\rangle$. For $W=\Phi_d(G)$, this gives
\begin{align}
W
 =\sum_{m=0}^{2d}\sum_{\gamma\in\mathcal I_m}
  \left(\sum_{i,j=1}^L
    G_{ij}\Gamma_{ij}^{(m,\gamma)}\right)
  O_\gamma^{(m)}.
\end{align}

All terms of degree greater than four vanish  when
\begin{align}
\sum_{i,j=1}^L
G_{ij}\Gamma_{ij}^{(m,\gamma)}=0
\qquad
\text{for every }m>4\text{ and }\gamma\in\mathcal I_m.
\end{align}
These are linear equations in the entries of $G$.
Writing $g=\operatorname{vec}(G)$, we collect them as
\begin{align}
A_{\mathrm{trunc}}(d) g=  0,
\end{align}
with one row for each high-degree operator
$O_\gamma^{(m)}$. The entries of that row are the corresponding
coefficients $\Gamma_{ij}^{(m,\gamma)}$, ordered consistently
with $\operatorname{vec}(G)$.

The nullspace of $A_{\mathrm{trunc}}(d)$ therefore contains the
Gram matrices whose terms of degree greater than four cancel.
To obtain SOS operators, we also require $G\succeq0$. If
$B_1,\ldots,B_K$ form a real basis for the Hermitian matrices
satisfying the nullspace equations, we can write
\begin{align}
G=\sum_{\ell=1}^K x_\ell B_\ell\succeq0,
\qquad x_\ell\in\mathbb R.
\end{align}
The resulting operators $W=\Phi_d(G)$ are degree-$d$ SOS
operators with ladder degree at most four.  Next we show that the nullspace conditions cover all possible witnesses generated from the cone $\Sigma_{d}$.

\subsection{Completeness of the two-body SOS nullspace construction}
First, we define the normal-ordered basis $\mathcal{O} = \{O_{\alpha}\}$ and partition it into two sets
\begin{align}
\mathcal{O}_{\leq 4} \;\;, \;\; \mathcal{O}_{>4}
\end{align}
corresponding to operator degree less than or equal to four or greater than four.  This partitioning allows us to define the projection of the coefficient onto the space for $\mathcal{O}_{>4}$ 
\begin{align}
\Pi_{>4} \colon \mathfrak{U}_{R} \mapsto \text{Span}\left(\mathcal{O}_{>4}\right).
\end{align}
The purpose of this projector is to use the equivalence between $\Pi_{>4}\left(\Phi_{d}(G)\right) = 0$ and $A_{\text{trunc}}(d)\text{vec}(G) = 0$.
\begin{theorem}\label{eq:sos_degree_d_completeness}
An at-most-two-body Hermitian operator $W$ has a degree-$d$ SOS representation if and only if there exists a positive semidefinite Gram matrix $G \succeq 0$ over the degree-$d$ generator basis such that $W = \vec{o}_{d}^{\dagger}G\vec{o}_{d}$ and $A_{\text{trunc}}(d)\text{vec}(G) = 0$. 
\begin{proof}
To prove this theorem we must demonstrate the equivalence $\Sigma_{d, \leq 2} = C_{d}$ where
\begin{align}
C_{d} \coloneqq& \left\{ \Phi_{d}(G) : G\succeq 0 , A_{\text{trunc}}(d)\text{vec}(G) = 0\right\} \\
\Sigma_{d, \leq 2} \coloneqq& \left\{ W \in \Sigma_{d} : \Pi_{>4}(W) = 0\right\}
\end{align}
which we will achieve by showing the forward statement $C_{d} \subseteq \Sigma_{d,\leq 2}$ and reverse statement $\Sigma_{d, \leq 2} \subseteq C_{d}$ are true. The forward statement checks if the degree-$d$ SOS can be represented as a gram matrix and satisfies the nullspace condition $A_{\text{trunc}}(d)\text{vec}(G) = 0$ always produces operators $W \in \Sigma_{d, \leq 2}$. 

Suppose $G\succeq 0$ and $ A_{\text{trunc}}(d)\text{vec}(G) = 0$. Let $W = \Phi_{d}(G) = \vec{o}_{d}^{\dagger}G\vec{o}_{d}$ then because $W = \Phi_{d}(G) \in \Sigma_{d}$ and the nullspace condition implies $\Pi_{>4}(W) =0$ then $W \in \Sigma_{d,\leq 2}$. 

Now, let $W \in \Sigma_{d, \leq 2}$ which means $W \in \Sigma_{d}$ and $\Pi_{>4}(W) = 0$. Because $W \in \Sigma_{d}$ then there is gram matrix representation $\Phi_{d}(G)$ for some $G\succeq 0$. $\Pi_{> 4}(W) = 0$ implies $A_{\text{trunc}}(d)\text{vec}(G) = 0$. Therefore, we do not miss any operators and $\Sigma_{d, \leq 2} \subseteq C_{d}$.
Combining the two inclusions proves the theorem.
\end{proof}
\end{theorem}

The nullspace construction captures every at-most-two-body operator admitting a degree-\(d\) SOS representation.

\begin{corollary}\label{cor:global_to_p_particle}
Let $W \in \Sigma_{d, \leq 2}$. Because $W$ is globally positive semi-definite across the entire Fock space, it is nonnegative on the restricted $p$-particle manifold ($\mathcal{H}_p$). Therefore, every $W \in \Sigma_{d, \leq 2}$ constitutes a valid $n$-representability witness for fixed-particle simulations. 
\end{corollary}

\begin{remark}
While Corollary \ref{cor:global_to_p_particle} guarantees that unweighted SOS operators are valid $n$-representability witnesses, the set $\Sigma_{d, \leq 2}$ is incomplete for fixed-particle simulations. There exist valid $p$-particle witnesses that possess negative eigenvalues on $N \neq p$ states, excluding them from the globally positive set $\Sigma_{d, \leq 2}$. Capturing these operators requires the transition to the weighted sum-of-squares (wSOS) framework discussed in Section~\ref{sec:weightedSOS}.
\end{remark}

\subsubsection{Symbolic construction of nullspace equations example}
The following section details the index mapping convention needed to implement the nullspace condition. This section can be skipped if practical implementation of the nullspace protocol is of no concern.
While a direct numerical approach in a pre-specified basis set size is feasible, our desire is to derive the generic structure for all witnesses at each SOS level. Thus, we return to the symbolic representation of the gram matrix and derive the nullspace conditions employing symbolic fermionic algebra software~\cite{liebenthal2025automated}. To achieve this we introduce a symbolic representation of operators in $\vec{o}$ where the ladder operator monomial in $\{a_{i},a_{i}^{\dagger},\mathbb{1}\}^{d}$ is indexed by $\sigma$ and $I = (i_{1}, i_{2}, ...)$ such that
\begin{align}
A_{\sigma}(I) = a_{i_{1}}^{\sigma_{1}}...a_{i_{m}}^{\sigma_{m}}
\end{align}
where $a_{i_k}^{\sigma_k=x} = a_{i_k}^\dagger$ and $a_{i_k}^{\sigma_k=o} = a_{i_k}$ 
which follows the convention from Ref.~\cite{PhysRevLett.108.263002}. An element of $\Sigma_{d}$ can, therefore, be constructed as
\begin{align}
W = \sum_{I,J,\sigma,\tau}G_{\sigma I, \tau J}A_{\sigma}(I)A_{\tau}(J)^{\dagger}.
\end{align}
We focus on particle conserving witnesses which imbues $G_{\sigma I, \tau J}$ with structure based on the number of fermionic raising and lowering operators in $A_{\sigma}(I)A_{\tau}(J)^{\dagger}$. If $\Delta N(\sigma) = n_{x} - n_{o}$ in $\sigma$, where $n_{x}$ is the number of $x$ characters in $\sigma$, then the particle conservation requirement is $\Delta N(\sigma) = \Delta N(\tau)$ (note that $\Delta N(\tau)$ is before taking the adjoint of $A_{\tau}(J)$). We can define $\eta$-particle level set
\begin{align}
\mathcal{S}_{\eta} = \{ \sigma  \in \{x, o\}^{d} : \Delta N(\sigma) = \eta\}
\end{align}
which allows us to write $W$ as 
\begin{align}
W = \sum_{\eta} \sum_{I, J, \sigma,\tau \in \mathcal{S}_{\eta}}G_{\sigma I, \tau J}^{(\eta)}A_{\sigma}(I)A_{\tau}(J)^{\dagger}
\end{align}
For example,
\begin{align}
\mathcal{S}_{+4} &={xxxx}, \\
\mathcal{S}_{+2} &={xxxo,xxox,xoxx,oxxx}, \\
\mathcal{S}_{0} &={xxoo,xoxo,xoox,oxxo,oxox,ooxx}, \\
\mathcal{S}_{-2} &={xooo,oxoo,ooxo,ooox}, \\
\mathcal{S}_{-4} &={oooo}
\end{align}
are all the index sets for $d=4$. Finally, we note that we need an index permutation mapping for the index set $(I, J)$ because $A_{\sigma}(I)A_{\tau}(J)^{\dagger}$ can be normal ordered into a label ordering that is not canonical. For example, consider rank-$8$ particle conserving operators
\begin{align}
E_{8}(y_{8}) &= a_{J}^{\dagger}a_{K}^{\dagger}a_{L}^{\dagger}a_{M}^{\dagger}a_{T}a_{R}a_{Q}a_{P} \\
y_{8} &\coloneqq ((J,K,L,M);(P, Q, R, T))
\end{align}
can be considered the canonical ordering for the 8-mode terms we want to ensure do not appear in our witness $W$. Next we introduce a map $P_{\sigma, \tau}(I, J)$ that permutes the indices of inputs $(I, J)$ according to the values of $\sigma$ and $\tau$ so that all symbols corresponding with $x$ are on the left and all symbols corresponding with $o$ are on the right. To see this more clearly we can consider the $d=4$ example again and define the index permutation $P_{xxxo,xxxo}((j, k, l, m);(p, q, r, t))$ as 
\begin{align}
P_{xxxo,xxxo}((j, k, l, m);(p, q, r, t)) = ((j, k, l, t);(p, q, r, m))
\end{align}
or a more general permutation $P_{xxxo,xxox}$
\begin{align}
P_{xxxo,xxox}((j, k, l, m);(p, q, r, t)) = ((j, k, l, r);(p, q, t, m)).
\end{align}
The permutation allows us to connect Gram matrix elements with a canonical ordering of ladder operators $y$. Let $q_{8}(y)$ be the coefficient associated with $E_{8}(y)$ then the contribution to $q_{8}(y)$ is 
\begin{align}
q_{8}(y) = \sum_{\eta}\sum_{\sigma,\tau \in \mathcal{S}_{\eta}}\Gamma_{\sigma,\tau}G_{\sigma,\tau}\left( P_{\sigma,\tau}^{-1}\left( y \right) \right)
\end{align}
where $P^{-1}_{\sigma,\tau}$ is the inverse index permutation and $\Gamma_{\sigma,\tau}$ is the coefficient obtained from normal ordering. The index permutation map $P$ takes a set number of input indices to the same number of output indices. This is insufficient to cover all cases from applying normal ordering to $A_{\sigma}(I)A_{\tau}(J)^{\dagger}$ due to the presence of contractions (Kronecker delta functions arising from the application of the anticommutation relations). Due to the antisymmetry in $q_{8}(y)$ we also account for all index swaps of the labels that produce the same canonically ordered operator. Let $\kappa$ label the contraction ordering (which indices are contracted) and let
\begin{align}
R_{\sigma,\tau,\kappa}^{(m)}(y, u)
\end{align}
be an index function that inserts the canonical indices $y$ and dummy contraction indices $u$ into the gram-tensor arguments. Thus,
\begin{align}
q_{m}(y) =& \sum_{\eta}\sum_{\sigma,\tau \in \mathcal{S}_{\eta}} \sum_{\kappa \in \mathcal{K}_{m}(\sigma,\tau)}\Gamma_{\sigma,\tau,\kappa}^{(m)}\sum_{\boldsymbol{u}}G_{\sigma,\tau}\left(R_{\sigma,\tau,\kappa}^{(m)}(y,\boldsymbol{u}) \right)\\
\coloneqq& \left[ A_{m}\text{vec}(G)\right]\left(y_{m}\right)
\end{align}
is the general form of the contribution to the coefficient tensor for $m$-mode ladder operators in canonical ordering $y$. The nullspace contraction requires that these coefficient matrices are equal to zero which provides a set of linear relationships the $G_{\sigma I, \tau J}$ gram-tensors must satisfy such that $W$ is a valid witness in $\Sigma_{d, \leq 2}$.
 We provide an example for constructing the symbolic nullspace condition in Appendix~\ref{app:contraction_condition_construction}. It should be noted that it is possible to extend the blocking structure in $G$ to include additional symmetries like $S_z$ relevant to electronic structure. The symmetry partitioning of $G$ with respect to $S_z$ is discussed in Appendix~\ref{app:gram_symmetries} and reproduces the block structure in Ref.~\cite{rubin2026near}.

\subsection{Nullspace conditions in the primal problem}\label{sec:nullspaceprimal_sos}
This section derives how the nullspace conditions should appear in the primal form of the SOS optimization and ultimately the conditions that the 2-RDM must satisfy to be representable. Specifically, the high-degree cancellation equations in the SOS program introduce auxiliary higher-degree moments in the primal program.
Upon conversion of the SOS to its dual, each linear equation produces
an unrestricted Lagrange multiplier that enters the lifted primal
moment matrix shaping the feasible space of RDMs.

Recall that the dual SOS is formulated as
\begin{align}
\max_{E_{\mathrm{SOS}},G}\quad
    &E_{\mathrm{SOS}}
    \\
\text{s.t.}\quad
    &H-E_{\mathrm{SOS}}\mathbb{1}
    = \vec{o}_d^{\dagger} G \vec{o}_d
    \nonumber \\
    &G\succeq0. \nonumber
\end{align}
This expression implicitly requires the coefficients of all
normal-ordered operators of degree greater than four to vanish. 
Now, let $A_{\leq4}g$ denote the coefficients of the canonical
degree-$0$, degree-$2$, and degree-$4$ operators, and let
$A_{>4}g$ denote the coefficients of all higher-degree
operators.  If $h$ is the coefficient vector of $H$ and $e_0$ selects
the identity operator, the SOS program can be written as
\begin{align}
\max_{E_{\mathrm{SOS}},G}\quad
    &E_{\mathrm{SOS}}
    \\
\text{s.t.}\quad
    &h-E_{\mathrm{SOS}}e_0
    =
    A_{\leq4}g,
    \label{eq:dual-low-body-match}
     \\
    & A_{>4}g=0,
    \label{eq:dual-high-body-null}
    \\
    &G\succeq0.
\end{align}
For quartic generators,
\begin{align}
     A_{>4}
    =
    A_6\oplus A_8,
\end{align}
where $ A_6g=q_6$ and $ A_8g=q_8$ are the coefficients for canonically ordered $a_{i}^{\dagger}a_{j}^{\dagger}a_{k}^{\dagger}a_{r}a_{q}a_{p}$ and $a_{i}^{\dagger}a_{j}^{\dagger}a_{k}^{\dagger}a_{l}^{\dagger}a_{s}a_{r}a_{q}a_{p}$ operators, respectively. 

Next we derive the dual of the SOS program by introducing a Lagrangian. First, we introduce a multiplier $x$ for
Eq.~\eqref{eq:dual-low-body-match} and a multiplier $u$ for
Eq.~\eqref{eq:dual-high-body-null}.  The Lagrangian is
\begin{align}
\mathcal L
={}&
E_{\mathrm{SOS}}
+
x^T
\left(
h-E_{\mathrm{SOS}}e_0- A_{\leq4}g
\right)
-u^T A_{>4}g
\nonumber\\
={}&
h^Tx
+
E_{\mathrm{SOS}}\left(1-e_0^Tx\right)
-
\left(
 A_{\leq4}^{T}x
+
 A_{>4}^{T}u
\right)^Tg .
\label{eq:nullspace-primal-lagrangian}
\end{align}
Next, we define the lifted moment matrix
\begin{align}
    M(x,u)
    =
    \operatorname{vec}^{-1}
    \left(
         A_{\leq4}^{T}x
        +
         A_{>4}^{T}u
    \right)
\end{align}
such that $M(x,u)$ is the matrix of pseudoexpectation values with the same operator basis as $G$--i.e. $M(x,u)_{a,b} = \langle o_{a}^{\dagger}o_{b}\rangle$. Recall, $A_{\leq 4}$ is a map from the space of matrices $G$ to coefficient tensors corresponding to pseudoexpectation values (reduced density matrix elements) of ladder degree at most 4 (many-body rank at most 2) corresponding to the scalar, $1$-RDM, and $2$-RDM.  Similarly, $A_{> 4}$ maps to coefficients of operators greater than four and thus $A_{>4}^{T}(u)$ populates the components of $M(x,u)$ corresponding to these higher many-body-rank operators.
Then the final term in
Eq.~\eqref{eq:nullspace-primal-lagrangian} is
$-\langle M(x,u),G\rangle$.  The supremum over
$E_{\mathrm{SOS}}$ is finite only if
\begin{align}
    e_0^Tx=1,
\end{align}
while the supremum over $G\succeq0$ is finite only if
\begin{align}
    M(x,u)\succeq0.
\end{align}
The corresponding primal program is therefore
\begin{align}
\min_{x,u}\quad
    &h^Tx
    \label{eq:nullspace-primal-general} \\
\text{s.t.}\quad
    &e_0^Tx=1,
    \nonumber \\
    &
    \operatorname{vec}^{-1}
    \left(
        A_{\leq4}^{T}x
        +
        A_{>4}^{T}u
    \right)
    \succeq0. \nonumber
\end{align}
We emphasize that the variables $u_6$ and $u_8$ are Lagrange multipliers
for the equations $q_6=0$ and $q_8=0$; they are not themselves
positive-semidefinite variables. Only the assembled moment matrix in
Eq.~\eqref{eq:nullspace-primal-general} is constrained to be positive
semidefinite. In Appendix~\ref{app:standard-rdm-adjoint-examples} we provide examples of the correspondence between $M(x,u)$ and traditional v2RDM non-negativity constraints along with a gram matrix interpretation of the $(2,4)$ conditions described in Ref.~\cite{PhysRevLett.108.263002}.

The auxiliary variable $u$ appearing in the primal form impacts the shape of allowed RDMs.
To see this, consider for any feasible Gram matrix $G$, with
$g=\operatorname{vec}(G)$, one has
\begin{align}
    \left\langle M(x,u),G\right\rangle
    &=
    x^{T}A_{\leq4}g
    +
    u^{T} A_{>4}g
    \nonumber\\
    &=
    x^{T} A_{\leq4}g.
\end{align}
The lifted variable therefore changes only the positive-semidefinite
representation of the primal cone.
The purpose of transforming to the primal representation is to derive conditions for a given 2-RDM to be certified representable up to the $\Sigma_{d, \leq 2}$ level. 

For a given set of RDMs, $x$, we can decompose $M(x, u)$ by linearity
\begin{align}\label{eq:m_xu_decomposition}
M(x, u) = M_{0}(x) + A^{*}(u)
\end{align}
where $M_{0}(x) = \text{vec}^{-1}(A^{T}_{\leq 4} x)$ and $A^{*}(u) = \text{vec}^{-1}(A^{T}_{>4}u)$. 
For a valid $G$, $M(x, u)$ can be written just as a function of $M_{0}(x)$
\begin{align}
\langle M(x, u), G\rangle = \langle M_{0}(x), G\rangle 
\end{align}
because $\langle A^{*}(u), G\rangle = 0$. What this shows is that the witness, $G$, cannot distinguish $M_{0}(x)$ from $M_{0}(x)+ A^{*}(u)$; there is an equivalence class $\left[ M_{0}(x) \right] = M_{0}(x) + \text{range}(A^{*})$ of lifted RDM solutions for the valid witness we are considering. The eigenvalues of the matrices in this equivalence class are not necessarily the same or even all non-negative--i.e. the same low-order data $x$ have equivalent high-order extensions, with respect to $G$. 
Although every matrix in this equivalence class has the same pairing with every admissible Gram matrix, these matrices need not all be positive semidefinite. We therefore search for a positive-semidefinite representative
\begin{align}\label{eq:dualcone_sos_nrepprimal}
\exists u : M_{0}(x) + A^{*}(u) \succeq 0 \implies x \in \tilde{\Sigma}_{d, \leq 2}
\end{align}
which can be transformed into a program to check the $n$-representability of a given $x$ up to $\Sigma_{d,\leq 2}$. Specifically,
\begin{align}
\text{find}\quad & u \label{eq:fixed-rdm-feasibility}\\
\text{subject to}\quad
& M_0(x)+A^*(u)\succeq0  \nonumber.
\end{align}
where $x$ is fixed; only the auxiliary higher-degree moments $u$ are varied.

If a feasible $u$ exists, then for every Gram matrix satisfying
$G\succeq0$ and $A_{>4}g=0$,
\begin{align*}
x^T A_{\leq4}g
&=\langle M_0(x),G\rangle\\
&=\langle M_0(x)+A^*(u),G\rangle
\geq0.
\end{align*}

\section{Weighted SOS Witnesses}\label{sec:weightedSOS}

While this structure is sufficient to represent positive polynomials of non-commuting variables~\cite{helton2002positive}, it does not naturally include algebraic constraints or symmetry ({\em e.g.}, particle number, spin, etc.) constraints.  Accounting for such constraints requires a generalized positive ansatz, which is known as a weighted sum of squares (wSOS)~\cite{helton2004positivstellensatz}. The weighted sum of squares features $\Sigma_{d}$ along with a congruence transformed set of constraints $q_{i} \geq 0$ 
\begin{align}
{}_{\text{w}}\Sigma_{d} \coloneqq \Bigg \{ B &= \sum_{i}R_{i}^{\dagger}R_{i} + \sum_{ji}x_{ij}^{\dagger}q_{j}x_{ij}\; \Bigg \vert \; R_{i} \in \text{Span}\left( \{a_{k}, a_{l}^{\dagger}, \mathbb{1}\}^{d}\right), \nonumber \\
& x_{ij} \in  \text{Span}\left( \{a_{k}, a_{l}^{\dagger}, \mathbb{1}\}^{d'}\right), d' = d - \left\lceil\max(\text{deg}(q_{j}))/2 \right\rceil \Bigg \}
\end{align}
where the polynomials of non-commutative operators $q_{i}$ define the non-negativity domain of interest. Of particular interest to the fermionic marginal problem is the constraint defining the particle number ($\eta$) manifold of Fock space. More broadly, spatial-, spin-, or algebraic-symmetry can also be encoded as a linear constraint. For example, the anticommutation relations can be included as constraints for the positivity domain--e.g. $\left\{a_{i}, a_{j}\right\} = 0$; though, as explained in Ref.~\cite{pironio2010convergent}, it is simpler to include equality constraints as linear maps from the appropriate gram matrix elements to Hamiltonian coefficients.  Taking the dual of the primal problem, one can show that the SOS certificate for the equality constraints, $r_{i}$, is of the form
\begin{align}
\sum_{i}f_{i}r_{i} + r_{i}^{\dagger}f_{i}^{\dagger}
\end{align}
where $f_{i}$ are degree $d'$ such that $\text{deg}(f_{i}r_{i}) \leq 2d$.  Thus, the weighted SOS for the $n$-representability problem is
\begin{align}\label{eq:wsos_nrep}
{}_{\text{wn}}\Sigma_{d}  \coloneqq \Bigg \{ B &= \sum_{i}R_{i}^{\dagger}R_{i} + \sum_{i}f_{i}r_{i} + r_{i}^{\dagger}f_{i}^{\dagger}\; \Bigg \vert \; R_{i} \in \text{Span}\left( \left\{a_{k}, a_{l}^{\dagger}, \mathbb{1}\right\}^{d}\right),  \nonumber \\
 &f_{i} \in  \text{Span}\left( \{a_{k}, a_{l}^{\dagger}, \mathbb{1}\}^{d'}\right), d' = 2d - \max(\text{deg}(r_{i}))\Bigg \} 
\end{align}
where one of the $r_{i}$ is $r = N - \eta \mathbb{1}$. Due to the extra non-negativity, or congruency constraint, the weighted SOS enjoys the inclusion relation
\begin{align}\label{eq:cone_subset}
\Sigma_{d}  \subseteq {}_{\text{wn}}\Sigma_{d}.
\end{align}
The weighted cone therefore permits witnesses that are
nonnegative on the target particle-number subspace without
requiring global positivity on Fock space. 

Next, we show that any Hermitian, particle-conserving operator
$W$, with $[W,N]=0$ and $W|_{\mathcal{H}_\eta}\succeq0$,
admits a weighted SOS representation at some sufficiently
large finite level. Since $W$ preserves each particle-number
sector, this follows by adding a number-variance term to make
$W$ globally positive semidefinite:
\begin{align}
S \coloneqq W + \tau\left(N - \eta \mathbb{1}\right)^{2} \succeq 0 \;,\;\; \tau \geq ||W||_{\text{op}}
\end{align}
which is automatically non-negative on the $\eta$-particle subspace $S|_{\mathcal{H}_{\eta}} = W|_{\mathcal{H}_{\eta}} \succeq 0$ and on any other subspace $\mu \neq \eta$
\begin{align}
S|_{\mathcal{H}_{\mu}} = W|_{\mathcal{H}_{\mu}}  + \tau(\mu - \eta)^{2} \mathbb{1} \succeq -||W||_{\text{op}}\mathbb{1} + \tau \mathbb{1} \succeq 0.
\end{align}
Because $S$ is now globally non-negative it has a factorization $S = R^{\dagger}R$ for some fermionic ladder operator polynomial $R$.  The added variance term can be written as the ideal by setting $f = -\frac{\tau}{2}r$ such that $f r + r f^{\dagger} = -\tau r^{2}$ which implies
\begin{align}
W = R^{\dagger}R + f r + r f^{\dagger}.
\end{align}
This establishes a weighted SOS representation at some
sufficiently large finite level but does not provide a degree bound for the SOS factor. In the next section we use the same nullspace construction to connect the exhaustive representation of ${}_{\text{wn}}\Sigma_{d}$ as the weighted SOS of degree-$d$.

\subsection{The Weighted SOS Nullspace}

In order to adapt our nullspace construction from the global SOS to the weighted SOS we construct the homogeneous linear equations for the high-body terms to include the weighting polynomial $f r + r f^{\dagger}$. 
Let $A_{\text{SOS}}$ be the truncation matrix mapping the Gram variables $\text{vec}(G)$ to the coefficients of high-body operators, and let $A_{\text{sym}}$ be the matrix mapping the coefficients $\vec{f}$ to those same high-body operators. Next, we define the subset of elements whose normal-ordered terms of ladder degree greater than four vanish
\begin{align}
{}_{\text{wn}}\Sigma_{d, \leq 2} \coloneqq \left\{ W \in {}_{\text{wn}}\Sigma_{d} : \Pi_{>4}(W) = 0, [W, N] = 0\right\}
\end{align}
which will be the space of witnesses of the $n$-representability problem built from degree-$d$ wSOS.
The requirement that the total witness operator $W \in {}_{\text{wn}}\Sigma_{d, \leq 2}$ is exactly 2-body implies that the sum of the high-body components from the SOS part and the shift part $(fr+rf^{\dagger})$ must cancel. This defines an augmented nullspace condition:
\begin{equation}
A_{\text{SOS}} \text{vec}(G) + A_{\text{sym}} \text{vec}(f) = \vec{0} \implies \begin{bmatrix} A_{\text{SOS}} & A_{\text{sym}} \end{bmatrix} \begin{pmatrix} \text{vec}(G) \\ \text{vec}(f) \end{pmatrix} = \vec{0}.
\end{equation}

We now show that the augmented nullspace construction captures
every at-most-two-body operator admitting a degree-$d$ weighted
SOS representation. This proof follows an identical pattern as the previous proof for completeness of $\Sigma_{d, \leq 2}$. 
Let $f_z$ parameterize the full space of allowed multipliers
$\mathcal{V}_{2d-2}$. Here $z\in\mathbb{R}^{K}$ collects the real
and imaginary parts of the multiplier coefficients, so every
allowed multiplier has a representation $f=f_z$.  Next, define
\begin{align}
\psi_d(G,z)
:=
\Phi_d(G)+f_zr+rf_z^\dagger.
\label{eq:weighted-gram-map}
\end{align}
The Gram-matrix representation of $\Sigma_d$ and the
parameterization of the multipliers give
\begin{align}
{}_{\text{wn}}\Sigma_{d}
=
\left\{
\psi_d(G,z):
G\succeq0,\ z\in\mathbb{R}^{K}
\right\}.
\label{eq:weighted-gram-cone}
\end{align}

The matrices $A_{\mathrm{SOS}}$ and $A_{\mathrm{sym}}$ give
the high-degree coefficients of $\Phi_d(G)$ and
$f_zr+rf_z^\dagger$, respectively, in the same canonical
normal-ordered basis. Because these basis operators are
linearly independent,
\begin{align}
\Pi_{>4}\bigl(\psi_d(G,z)\bigr)=0
\quad\Longleftrightarrow\quad
A_{\mathrm{SOS}}\operatorname{vec}(G)
+
A_{\mathrm{sym}}z=0.
\label{eq:weighted-projection-nullspace}
\end{align}
Thus the absence of high-degree terms is equivalent to
membership of the combined parameter vector in the augmented
nullspace:
\begin{align}
\begin{pmatrix}
\operatorname{vec}(G)\\
z
\end{pmatrix}
\in
\ker
\begin{pmatrix}
A_{\mathrm{SOS}} & A_{\mathrm{sym}}
\end{pmatrix}.
\label{eq:weighted-augmented-kernel}
\end{align}

\begin{theorem}[Completeness of the weighted SOS nullspace construction]
\label{thm:weighted-nullspace-completeness}
An at-most-two-body Hermitian operator $W$, where $\left[W, N\right]=0$, belongs to
${}_{\operatorname{wn}}\Sigma_{d, \leq 2}$ if and only if there exist
$G\succeq0$ and $z\in\mathbb{R}^{K}$ such that
\begin{align}
W &= \psi_d(G,z),\\
A_{\mathrm{SOS}}\operatorname{vec}(G)
+
A_{\mathrm{sym}}z &= 0.
\end{align}
\end{theorem}

\begin{proof}
To prove the theorem, we demonstrate the equality
${}_{\operatorname{wn}}\Sigma_{d,\leq2}=\mathcal{C}_d^{\mathrm{wn}}$,
where
\begin{align}
{}_{\operatorname{wn}}\Sigma_{d,\leq2}
&:=
\left\{
W\in {}_{\operatorname{wn}}\Sigma_d:
\Pi_{>4}(W)=0, [W, N] = 0
\right\},\\
\mathcal{C}_d^{\mathrm{wn}}
&:=
\left\{
\psi_d(G,z):
\begin{array}{l}
G\succeq0,\ z\in\mathbb{R}^{K},\\
A_{\mathrm{SOS}}\operatorname{vec}(G)
+A_{\mathrm{sym}}z=0, \\
\left[\psi_d(G, z), N\right] = 0
\end{array}
\right\}.
\end{align}
We show the forward inclusion
$\mathcal{C}_d^{\mathrm{wn}}
\subseteq {}_{\text{wn}}\Sigma_{d,\leq2}$
and the reverse inclusion
${}_{\text{wn}}\Sigma_{d,\leq2}
\subseteq\mathcal{C}_d^{\mathrm{wn}}$.

Every feasible augmented-nullspace pair produces a
degree-$d$ two-body weighted SOS witness. To show this, let $W\in\mathcal{C}_d^{\mathrm{wn}}$. Then there exist
$G\succeq0$ and an allowed multiplier $f_z$ such that
\begin{align}
W=\Phi_d(G)+f_zr+rf_z^\dagger
\end{align}
and
\begin{align}
A_{\mathrm{SOS}}\operatorname{vec}(G)
+
A_{\mathrm{sym}}z=0.
\end{align}
Because $G\succeq0$, we have $\Phi_d(G)\in\Sigma_d$.
Adding the allowed multiplier term therefore gives
$W\in {}_{\operatorname{wn}}\Sigma_d$. The augmented nullspace condition
implies $\Pi_{>4}(W)=0$, so
$W\in {}_{\operatorname{wn}}\Sigma_{d,\leq2}$. Consequently,
\begin{align}
\mathcal{C}_d^{\mathrm{wn}}
\subseteq
{}_{\text{wn}}\Sigma_{d,\leq2}.
\end{align}

Every degree-$d$ two-body weighted SOS witness has a
feasible augmented-nullspace pair. To show this,
let $W\in {}_{\text{wn}}\Sigma_{d,\leq2}$. By definition,
$W\in {}_{\operatorname{wn}}\Sigma_d$ and $\Pi_{>4}(W)=0$. Hence there
exist $S\in\Sigma_d$ and an allowed multiplier
$f\in\mathcal{V}_{2d-2}$ such that
\begin{align}
W=S+fr+rf^\dagger.
\end{align}
The Gram-matrix representation of $\Sigma_d$ gives some
$G\succeq0$ with $S=\Phi_d(G)$. Since $f_z$ spans the full
allowed multiplier space, there is also some $z$ with
$f=f_z$. Therefore,
\begin{align}
W=\psi_d(G,z).
\end{align}
Using $\Pi_{>4}(W)=0$ in
Eq.~\eqref{eq:weighted-projection-nullspace} gives
\begin{align}
A_{\mathrm{SOS}}\operatorname{vec}(G)
+
A_{\mathrm{sym}}z=0.
\end{align}
Thus $(G,z)$ is a feasible augmented-nullspace pair and
$W\in\mathcal{C}_d^{\mathrm{wn}}$. Consequently,
\begin{align}
{}_{\text{wn}}\Sigma_{d,\leq2}
\subseteq
\mathcal{C}_d^{\mathrm{wn}}.
\end{align}
Combining the two inclusions proves the theorem.
\end{proof}

The augmented nullspace construction therefore captures every
at-most-two-body element of the prescribed degree-$d$ weighted
SOS cone. Importantly, the high-degree terms of $\Phi_d(G)$
and $f_zr+rf_z^\dagger$ need not vanish separately; only their
sum must vanish. The positive-semidefinite constraint remains
on $G$, while the multiplier coordinates remain unrestricted.

The new free weighting polynomial $f$ along with the ideal constraint translate into a new equality constraint when transforming the dual program to the primal program. 
The SDP for the weighted SOS optimization has the form
\begin{align}
\max_{E_{\text{SOS}}, G, z} \;E_{\text{SOS}} \\
\text{s.t.} \; h - E_{\text{SOS}} e_{0} = A_{\leq 4} g + B_{\leq 4} z \nonumber \\
A_{>4}g + B_{>4}z = 0 \nonumber \\
G\succeq 0, z \in \mathbb{R}^{K} \nonumber
\end{align}
where $B_{\leq 4}$ and $B_{>4}$ are the maps that take $z$ to the low- and high-degree canonical ordered coefficient tensors in $f_{z}r + rf_{z}^{\dagger}$. We also note that in this notation $A_{>4} = A_{\text{SOS}}$ and $B_{>4} = A_{\text{sym}}$. The dual of this program can be obtained just as in Section~\ref{sec:nullspaceprimal_sos} and has the form
\begin{align}
\min_{x, u} h^{T}x \\
\text{s.t.} \; e_{0}^{T}x = 1, \nonumber \\
\text{vec}^{-1}\left(A_{\leq 4}^{T}x + A_{>4}^{T}u\right) \succeq 0, \nonumber \\ 
B_{\leq 4}^{T}x + B_{>4}^{T}u = 0 \nonumber 
\end{align}
where $u$ is a free multiplier variable for the constraint that the higher degree ladder operators evaluate to zero $-u^{T}\left(A_{>4}g + B_{>4}z\right)$.

Just as in the unweighted case we can derive the conditions on a given set of RDMs to be representable up to the ${}_{\text{wn}}\Sigma_{d, \leq 2}$ level. Define the lifted moment matrix and the symmetry-constraint
residual by
\begin{align}
M(x,u)
&=\operatorname{vec}^{-1}
\bigl(A_{\leq4}^Tx+A_{>4}^Tu\bigr),
\label{eq:weighted-moment-matrix}\\
c(x,u)
&=B_{\leq4}^Tx+B_{>4}^Tu.
\label{eq:weighted-symmetry-residual}
\end{align}
Just as in Eq.~\eqref{eq:m_xu_decomposition} we can represent $M(x,u) = M_{0}(x) + A^{*}(u)$ which along with the additional constraint $c(x,u)=0$ parameterizes feasible truncated moment extensions to verify a given set of RDMs is representable up to ${}_{\text{wn}}\Sigma_{d, \leq 2}$
\begin{align}\label{eq:weighted_primal_nrep}
\text{find} \;& u \\
\text{s.t.} \;& M_{0}(x) + A^{*}(u) \succeq 0 \nonumber \\
              & B_{\leq4}^{T}x+B_{>4}^{T}u=0.\nonumber
\end{align}
where $u$ does not have any positivity constraints. Just as in Eq.~\eqref{eq:dualcone_sos_nrepprimal} finding a feasible solution to the program in Eq.~\eqref{eq:weighted_primal_nrep} verifies $x \in {}_{\text{wn}}\tilde{\Sigma}_{d,\leq 2}$. In other words, $x$ is a set of RDMs that is representable up to witnesses generated by ${}_{\text{wn}}\Sigma_{d, \leq 2}$.

\subsection{Complete physical witnesses and $n$-representability}\label{sec:complete_nrep_phys_witness}
Finally, we can complete the construction of all witnesses that characterize a given $x$ is $n$-representable. 
Fix $2\leq\eta\leq R$ and $r=N-\eta\mathbb{1}$. 
The physical witness cone is
\begin{align}
\mathcal{W}_{\eta,\leq2}
:=
\left\{
W=W^\dagger:
[W,N]=0,\ \deg W\leq4,\
W|_{\mathcal{H}_\eta}\succeq0
\right\}.
\end{align}
These operators need only be nonnegative on
$\mathcal{H}_\eta$, not on the entire Fock space. Because the degree cutoffs are inclusive, every certificate
available at a lower level is also available at a higher level:
\begin{align}
{}_{\text{wn}}\Sigma_{d,\leq2}
\subseteq
{}_{\text{wn}}\Sigma_{\eta,\leq2},
\qquad 2\leq d\leq\eta.
\end{align}
We now show that the terminal cone contains every physical
two-body witness.

\begin{theorem}[Completeness of physical witnesses]
The complete weighted SOS hierarchy satisfies
\begin{align}
\mathcal{W}_{\eta,\leq2}
=
{}_{\operatorname{wn}}\Sigma_{\eta,\leq2}.
\end{align}
Thus $d=\eta$ is sufficient to generate all physical two-body
witnesses, including those available at lower levels.
\end{theorem}

\begin{proof}
Our goal is to show that physical positivity on
$\mathcal{H}_\eta$ is equivalent to an allowed weighted SOS
representation at $d=\eta$. The forward construction has two parts: 1) we use the
positive matrix $W|_{\mathcal{H}_\eta}$ as a Gram matrix to
construct $S_\eta\in\Sigma_\eta$ with the same action as $W$
on the target sector and 2) we must also show that their difference on
Fock space is an allowed particle-number shift. We cannot stop at (1) because positivity on the target sector does not imply a weighted SOS certificate. To construct the shift obtained from their difference, we rewrite the scalar, one-body,
and two-body terms of $W$ as purely $\eta$-body terms,
recording an explicit multiple of $r$ at each replacement.
We use this lifting because the constructed $S_\eta$ is purely
$\eta$-body, and the coefficients of a purely $\eta$-body
operator are determined by its matrix on
$\mathcal{H}_\eta$. This will identify the final lifted
operator with $S_\eta$.

The other inclusion is immediate. If
$W=S+fr+rf^\dagger$ is a weighted witness, then for any
$|\psi\rangle\in\mathcal{H}_\eta$,
\begin{align}
\langle\psi|W|\psi\rangle
=
\langle\psi|S|\psi\rangle
\geq0,
\end{align}
because $r|\psi\rangle=0$ and $\langle\psi|r=0$.
Therefore
${}_{\text{wn}}\Sigma_{\eta,\leq2}
\subseteq\mathcal{W}_{\eta,\leq2}$.
For the reverse inclusion, fix
$W\in\mathcal{W}_{\eta,\leq2}$.

To start we construct an SOS agreeing with $W$ on the target sector.
For each increasing $\eta$-tuple
$I=(i_1<\cdots<i_\eta)$, define
\begin{align}
d_I=a_{i_\eta}\cdots a_{i_1}\in\mathcal{V}_\eta,
\qquad
|I\rangle=d_I^\dagger|0\rangle.
\end{align}
These annihilation operators satisfy
$d_I|J\rangle=\delta_{IJ}|0\rangle$. Consequently,
\begin{align}
\left.d_I^\dagger d_J\right|_{\mathcal{H}_\eta}
=
|I\rangle\langle J|.
\label{eq:physical_matrix_units}
\end{align}
Their products therefore resolve the full operator space
on the target sector.

Set
\begin{align}
G_{IJ}:=\langle I|W|J\rangle,
\qquad
S_\eta:=\sum_{I,J}G_{IJ}d_I^\dagger d_J.
\end{align}
The physical-witness condition implies $G\succeq0$.
Factoring $G=C^\dagger C$ gives
\begin{align}
S_\eta
&=\sum_\alpha R_\alpha^\dagger R_\alpha,\\
R_\alpha
&=\sum_J C_{\alpha J}d_J\in\mathcal{V}_\eta,
\end{align}
so $S_\eta\in\Sigma_\eta$. Equation~\eqref{eq:physical_matrix_units}
also gives
\begin{align}
S_\eta|_{\mathcal{H}_\eta}
=
W|_{\mathcal{H}_\eta}.
\end{align}

Both $W$ and $S_\eta$ are operators on the entire Fock space.
The equality just established concerns only their action
on $\mathcal{H}_\eta$. It does not assert $W=S_\eta$ on
other particle-number sectors. We must now account for
their difference and verify it has the form $f r + rf^{\dagger}$ where $\deg f \leq 2\eta - 2$.

To show their difference is the desired symmetry shift multiplier we lift a lower-body term and record its correction.
Write the original witness as
\begin{align}
W=w_0\mathbb{1}+W_1+W_2,
\end{align}
where $W_k$ is its normal-ordered $k$-body part. 
The purpose of lifting is to rewrite these lower-body terms
in the same purely $\eta$-body form as $S_\eta$, without
changing the full operator $W$. Consider first
\begin{align}
W_1=\sum_{ij}h_{ij}a_i^\dagger a_j,
\qquad h_{ij}=h_{ji}^*.
\end{align}
Normal ordering its product with $N$ gives
\begin{align}
W_1N
=
W_1+\sum_{ijs}h_{ij}
a_i^\dagger a_s^\dagger a_s a_j.
\end{align}
Since $N-\mathbb{1}=(\eta-1)\mathbb{1}+r$ (See Eq.~\eqref{eq:number-normal-order} for a more thorough explanation), defining the
purely two-body replacement
\begin{align}
B_2:=
\frac{1}{\eta-1}
\sum_{ijs}h_{ij}a_i^\dagger a_s^\dagger a_s a_j
\end{align}
yields the exact identity
\begin{align}
W_1=B_2-\frac{W_1}{\eta-1}r.
\label{eq:physical_one_body_lift}
\end{align}
Thus the one-body term has been replaced by a two-body
term, with its exact correction displayed as a multiple
of $r$. The replacement agrees with the original term
on $\mathcal{H}_\eta$, while the correction preserves
the equality on the entire Fock space.

The same operation applies to any purely $k$-body operator
$Q_k$ with $k<\eta$. In normal ordering $Q_kN$, the
contractions reproduce $kQ_k$, and the uncontracted terms
are purely $(k+1)$-body. Therefore
\begin{align}
Q_{k+1}
:=
\frac{Q_k(N-k\mathbb{1})}{\eta-k}
\end{align}
is purely $(k+1)$-body, and
\begin{align}
Q_k
=
Q_{k+1}-\frac{Q_k}{\eta-k}r,
\qquad 0\leq k<\eta.
\label{eq:physical_body_lift}
\end{align}
For the scalar contribution, this starts with
\begin{align}
w_0\mathbb{1}
=
\frac{w_0}{\eta}N-\frac{w_0}{\eta}r.
\end{align}

Now we repeat the lifting and collect the multiplier terms.
Apply Eq.~\eqref{eq:physical_body_lift} to each contribution
of $W$ until its replacement reaches body order $\eta$.
At each stage, retain the correction proportional to $r$
and continue lifting only the replacement term. For example,
the replacement $B_2$ in
Eq.~\eqref{eq:physical_one_body_lift} is lifted again when
$\eta>2$, while the already-recorded correction
$-W_1r/(\eta-1)$ remains unchanged. Next, collect all final $\eta$-body replacements into an operator
$T_\eta$, and all correction factors into $A$. The result is
\begin{align}
W=T_\eta+Ar,
\label{eq:physical_lifted_decomposition}
\end{align}
as an exact identity on Fock space. At this stage,
$T_\eta$ is simply the result of rewriting $W$.

Because the highest-body term entering $A$ is $Q_{\eta-1}$ (from the final lifting step $k = \eta - 1$), $\deg A \le 2\eta - 2$.
Moreover, each replacement preserves Hermiticity and
particle conservation because $Q_k$ commutes with $N$, so
$Q_k(N-k\mathbb{1})$ is Hermitian whenever $Q_k$ is.
Starting from the Hermitian parts of $W$, we consequently
obtain
\begin{align}
A=A^\dagger,\qquad [A,N]=0.
\end{align}
The last step is to identify the lifted operator with the constructed SOS.
Both $T_\eta$ and $S_\eta$ are purely $\eta$-body. We can
therefore expand $T_\eta$ in the same normal-ordered basis:
\begin{align}
T_\eta=\sum_{I,J}t_{IJ}d_I^\dagger d_J.
\end{align}
The matrix-unit property
in Eq.~\eqref{eq:physical_matrix_units} implies
\begin{align}
\langle K|T_\eta|L\rangle=t_{KL}.
\end{align}
On the other hand, taking matrix elements of
Eq.~\eqref{eq:physical_lifted_decomposition} between
$\eta$-particle determinants gives
\begin{align}
\langle K|W|L\rangle
&=\langle K|T_\eta|L\rangle
  +\langle K|Ar|L\rangle\\
&=\langle K|T_\eta|L\rangle,
\end{align}
because $r|L\rangle=0$. Hence
\begin{align}
t_{KL}=\langle K|W|L\rangle=G_{KL}.
\end{align}
All coefficients in the two ladder-operator expansions
are identical, so
\begin{align}
T_\eta
=
\sum_{I,J}G_{IJ}d_I^\dagger d_J
=
S_\eta.
\end{align}
This is an equality between $T_\eta$ and $S_\eta$ as
Fock space operators.

For the final step, we make the substitution for $A$ as the polynomial $f$. Substituting into
Eq.~\eqref{eq:physical_lifted_decomposition} and setting
$f=A/2$ gives
\begin{align}
W&=S_\eta+fr+rf^\dagger,\\
S_\eta&\in\Sigma_\eta,\qquad
f\in\mathcal{V}_{2\eta-2}.
\end{align}
Here $fr+rf^\dagger=Ar$ because $A$ is Hermitian and
commutes with $r$. The total operator is the original
two-body witness $W$, so its terms of ladder degree
greater than four cancel exactly. Therefore
$W\in{}_{\text{wn}}\Sigma_{\eta,\leq2}$,
which proves the reverse inclusion.
\end{proof}
For normalized candidate low-order moments $x$, the feasibility problem in Eq.~\eqref{eq:weighted_primal_nrep} at $d=\eta$ is necessary and sufficient for ensemble $\eta$-representability.

\section{Complete 3-representability}
In this section we provide a numerical verification of the full solution we just derived by computing a lower bound certificate that is equivalent to the true ground state in a $3$-particle sector of a Richardson Hamiltonian. 
\subsection{Particle-number weights for degree-6 wSOS}
\label{sec:particle-number-weights}

For $d=3$, Eq.~\eqref{eq:wsos_nrep} permits a particle-number
multiplier of degree at most four. 
Within the particle-conserving ansatz, $[f, N]=0$, so
$fr+rf^\dagger=(f+f^\dagger)r$.
We may therefore take $f=f^\dagger$
and write
\begin{align}
f &= f_0\mathbb{1}
 + \sum_{ij}F^{(2)}_{i;j}E_2(i;j)
 + \sum_{ijkl}F^{(4)}_{ij;kl}E_4(ij;kl),
\label{eq:number-multiplier}
\end{align}
where
$E_{2k}(I;J)=
a_{i_1}^\dagger\cdots a_{i_k}^\dagger
a_{j_k}\cdots a_{j_1}$. We set $f$ to be a canonically ordered polynomial because we can use CAR to write any particle conserving polynomial of degree $d \leq 4$ in canonical ordering. 
All index sums are unrestricted, with no factorial prefactors.
Hermiticity is ensured by $f_0\in\mathbb R$,
$F^{(2)}_{i;j}=F^{(2)*}_{j;i}$, and
$F^{(4)}_{ij;kl}=F^{(4)*}_{kl;ij}$.

The CAR give the compact normal-ordering identity
\begin{align}
E_{2k}(I;J) N
= kE_{2k}(I;J)
 + \sum_s E_{2k+2}(Is;Js),
\qquad k=0,1,2,
\label{eq:number-normal-order}
\end{align}
where $Is$ denotes appending $s$ to $I$.
The first term collects the $k$ contractions; the second adds
the two number-operator slots. Consequently, the canonical
coefficient tensors of $W_N(f)=fr+rf^\dagger=2fr$ are
\begin{align}
q_0 &= -2\eta f_0, \nonumber\\
q_2(i;j)
 &= 2f_0\delta_{ij}+2(1-\eta)F^{(2)}_{i;j}, \nonumber\\
q_4(ij;kl)
 &= 2\mathcal A_2\!\left[
 F^{(2)}_{i;k}\delta_{jl}
 +(2-\eta)F^{(4)}_{ij;kl}\right], \nonumber\\
q_6(ijk;lmn)
 &= 2\mathcal A_3\!\left[
 F^{(4)}_{ij;lm}\delta_{kn}\right].
\label{eq:number-weight-maps}
\end{align}
Here $\mathcal A_k$ antisymmetrizes the creator and annihilator
tuples independently, with normalization $1/(k!)^2$.
These tensors multiply sums over all ordered tuples.
The same canonical collection must therefore be applied to both
the Gram and multiplier contributions.

Let $z$ collect the independent real coordinates of the Hermitian
multiplier coefficients, and let $B_m z$ denote their
contribution in the same coefficient basis as $A_m g$.
For a two-body Hamiltonian, the weighted SDP is
\begin{align}
\max_{E_{\mathrm{SOS}},G,z}\quad &E_{\mathrm{SOS}}
\nonumber\\
\text{s.t.}\quad
h-E_{\mathrm{SOS}}e_0
 &=  A_{\leq4}g+ B_{\leq4}z,
\nonumber\\
0 &= A_6g+ B_6z,
\qquad G\succeq0,
\label{eq:number-weighted-sdp}
\end{align}
with $z$ unrestricted. In particular, the sextic SOS and multiplier
coefficients cancel jointly; neither contribution must vanish
separately.

For an SDP expressed entirely with PSD variables, each free
Hermitian coefficient matrix can be represented as
\begin{align}
F^{(q)}=P_q-M_q,\qquad P_q,M_q\succeq0,
\qquad q=0,2,4,
\end{align}
where $F^{(0)}=(f_0)$ and the matrix dimensions are $1$, $R$,
and $R^2$, respectively. The four-index matrix uses rows $(i,j)$
and columns $(k,l)$. Thus six PSD blocks represent the free
multiplier exactly; their differences are the coefficients of
$f$ itself, retaining the factor of two in
Eq.~\eqref{eq:number-weight-maps}.

\subsection{The Richardson pairing Hamiltonian}\label{sec:ayers}

\begin{figure}[H]
    \centering
    \includegraphics[width=8cm]{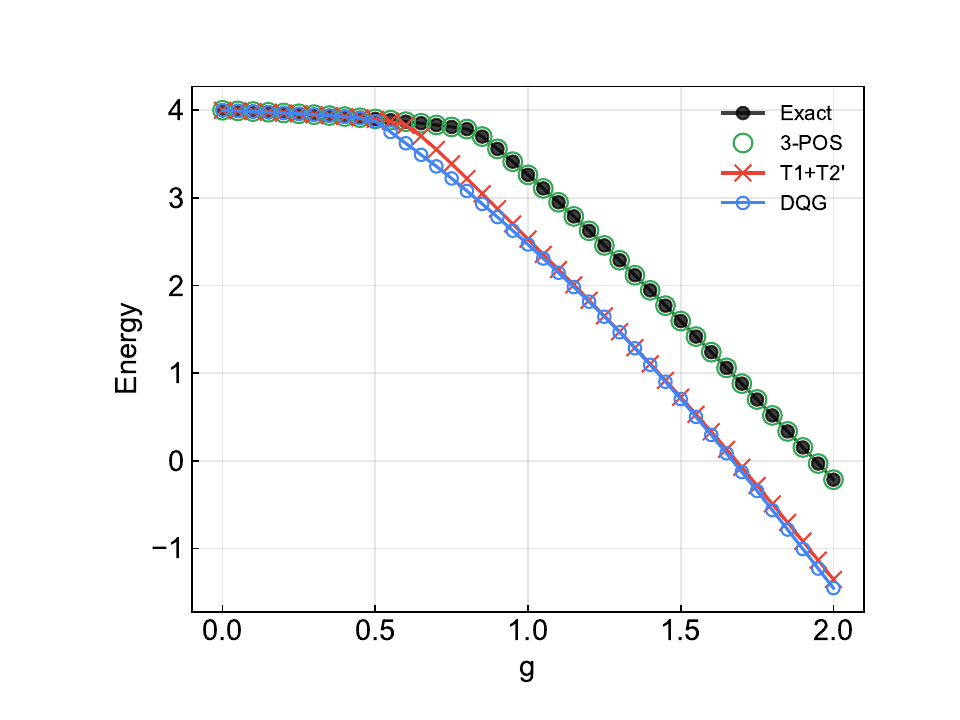}
    \includegraphics[width=8cm]{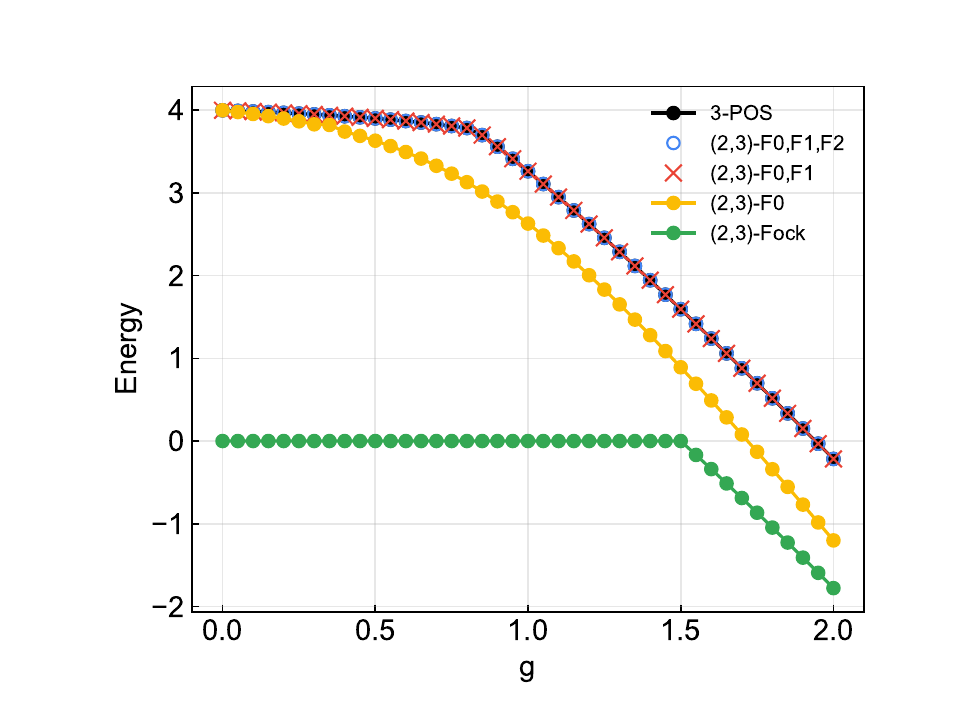}
    \caption{\textit{left}) Comparison of the exact ground state energy for spinless Richardson Hamiltonian (Eq.~\eqref{eq:ayers-H}) in the $3$-electron subspace as a function of interaction strength $g$ and various v2RDM methods. $3$-POS corresponds with $\Sigma_{3}$ witnesses and is exact for $3$-electron systems as we are optimizing over the full $3$-electron density matrix. $\mathrm{DQG}$ corresponds to $\Sigma_{2}$ while $\mathrm{T1}$+$\mathrm{T2'}$ corresponds to subcone of $\Sigma_{3, \leq 2}$ conditions constructed by anticommutation relation cancellation alone. Here $\Sigma_2$ and $\Sigma_3$ identify the SOS positivity
structure; the fixed-three-electron RDM calculations also
include the particle-number and RDM consistency constraints. \textit{right}) Comparison of the exact ground state energy for the Hamiltonian in Eq.~\eqref{eq:ayers-H} compared with the SOS dual methods described by the cone ${}_{\text{wn}}\Sigma_{d,\leq 2}$ with multipliers of degree-0 (constant denoted $\text{F0}$), degree-2 (constant and one-body denoted $\text{F0, F1}$), and degree-4 (constant, one-body, and two-body denoted $\text{F0, F1, F2}$) multiplier polynomials. Section~\ref{sec:degree-two-multipliers} proves that the degree-at-most-two and degree-at-most-four multiplier formulations have identical supremal energy bounds when the complete cubic SOS generator space is used.}
    \label{fig:dual_multiplier_degrees}
\end{figure}

We use a Richardson-type pairing Hamiltonian for spinless
fermions to compare conventional v2RDM relaxations with the
complete degree-three weighted SOS formulation\cite{Ayers_Unpublished}. The Hamiltonian is defined over eight modes with a diagonal one-body operator and where pair orbitals have the same energy monotonically increasing with mode index.  As such, the modes come in pairs where the partner of mode $i$ is $\bar{\imath} = i+1$ for odd $i$ and $\bar{\imath} = i-1$ for even $i$. Using the parameters
\begin{align}
\varepsilon_1 = \varepsilon_2 = 1, \quad \varepsilon_3 = \varepsilon_4 = 2, \quad \varepsilon_5 = \varepsilon_6 = 3, \quad \varepsilon_7 = \varepsilon_8 = 4 ,
\label{eq:ayers-eps}
\\
x_1 = -x_2 = \frac{1}{2\sqrt{15}}, \quad x_3 = -x_4 = \frac{1}{\sqrt{15}}, \quad x_5 = -x_6 = \frac{3}{2\sqrt{15}}, \quad x_7 = -x_8 = \frac{2}{\sqrt{15}} ,
\label{eq:ayers-x}
\end{align}
we use the Hamiltonian parameterized by $g$
\begin{align}
H(g) &= \sum_{i=1}^{8} \varepsilon_i\, a_i^{\dagger}a_i \;-\; 2g \sum_{i,j=1}^{8} x_i x_j\, a_i^{\dagger} a_{\bar{\imath}}^{\dagger} a_{\bar{\jmath}} a_j ,
\label{eq:ayers-H}
\end{align}
which we study in the three-particle sector. We computed lower bounds for various instantiations of programs defined with ${}_{\operatorname{wn}}\Sigma_{3,\leq 2}$ and differing multiplier degrees along with standard v2RDM calculations involving 2-positivity ($\mathrm{DQG}$) and partial 3-positivity ($\mathrm{DQGT_{1}T_{2}'}$).
The cubic positivity conditions of $\mathrm{DQGT_1T_2'}$ use the prescribed $\mathrm{T_1}$ and $\mathrm{T_2}'$ Gram block pairings to cancel three-body contributions. We have not included the additional lifting conditions of Ref.~\cite{PhysRevA.85.062507} which may further improve the energy of the primal lower bound SDP. In contrast, the complete weighted SOS formulation searches over all positive-semidefinite Gram matrices on
$\mathcal{V}_3$ and all allowed particle-number multipliers
of degree at most four, subject to the symmetry and
coefficient-matching constraints. The sextic SOS and
multiplier contributions cancel jointly, as specified
in Eq.~\eqref{eq:number-weighted-sdp}.
In Figure~\ref{fig:dual_multiplier_degrees} (left) the v2RDM lower bounds for a range of $g \in [0,2]$ are compared to the exact solution. We also plot $3$-positivity which is exact for a $3$-particle system because it can be made equivalent to exact diagonalization. We also compare to $\mathrm{DQG}$ to demonstrate that the $\mathrm{T_{1}}$ and $\mathrm{T_{2}'}$ conditions add little improvement in the lower bound energy certificate for this particular system. Both $\mathrm{DQG}$ and $\mathrm{DQGT_{1}T_{2}'}$ conditions constrain the particle count by constraining the $1$-RDM trace to be $3$ and the trace of the $2$-RDM is set to $6$. The $\mathrm{T_1}$ and $\mathrm{T_2}'$ metric matrices are expressed in terms
of the same one- and two-body moments through their CAR
identities; these calculations do not retain an independent
$3$-RDM. All calculations used SDPs generated with a locally modified version of \texttt{p$^\dagger$q}\cite{liebenthal2025automated}. The SDPs were solved with solvers implemented in a locally modified version of the  libsdp\cite{deprince_libsdp} library. We selected the stopping condition to be $1\times 10^{-5}$ Frobenius norm error on the primal residual equation $AX - b = 0$.

In the second (right) plot of Figure~\ref{fig:dual_multiplier_degrees} we show the dual SOS SDP with ${}_{\text{wn}}\Sigma_{3,\leq 2}$ and $\Sigma_{3, \leq 2}$ constraints. The green curve is the SOS with no multiplier and is equivalent to using the Gram cone $\Sigma_{3, \leq 2}$. This curve corresponds to the Fock space lower bound due to having no particle number restriction. To verify the performance of higher-degree multiplier polynomials we plot SOS dual program lower bound certificates generated by ${}_{\text{wn}}\Sigma_{3, \leq 2}$ using multiplier polynomials of degree-0, -$2$, or -$4$. While degree-$0$ (scalar) improves over the Fock space by allowing for a scalar multiple of the particle-number residual to reinforce $\Sigma_{3, \leq 2}$ it does not provide a substantial improvement over v2RDM $\mathrm{DQGT_{1}T_{2}'}$.  Multipliers of degree-$2$ and degree-$4$ allow cancellation with the degree-4 and degree-6 gram matrix elements of $\Sigma_{d,\leq 2}$ and recover the exact energy. Using Eq.~\eqref{eq:number-normal-order} we can see what the higher-degree multiplier polynomials can correspond to in the primal picture. Let $\mathcal{L}_{x,u}$ denote the linear functional recovering the RDM elements such that ${}^{k}D_{J}^{I} = \mathcal{L}_{x,u}\left(E_{2k}(I, J)\right)$ defines the $k$-RDM elements indexed by composite indices $(I, J)$. The particle number constraint $r$ implies
\begin{align}
\mathcal{L}_{x,u}\left(E_{2k}(I, J)r\right) = 0
\end{align}
which when substituting the normal-order identity of Eq.~\eqref{eq:number-normal-order} gives 
\begin{align}
\left(k - \eta\right){}^{k}D^{I}_{J} + \sum_{s}{}^{k+1}D_{Js}^{Is} = 0,\quad k=0,1,2
\end{align}
corresponding to the trace condition on the $1$-RDM, consistency between the $2$-RDM to the $1$-RDM, and consistency between the $3$-RDM and the $2$-RDM
\begin{align}
\sum_{s} {}^{1}D_{s}^{s} =& \eta \\
\sum_{s} {}^{2}D_{js}^{is} =& \left(\eta - 1\right){}^{1}D_{j}^{i} \\
\sum_{s} {}^{3}D_{kls}^{ijs} =& \left(\eta - 2\right){}^{2}D_{kl}^{ij} \label{eq:119}.
\end{align}
This translation reveals that the multiplier polynomial maps in Eq.~\eqref{eq:weighted_primal_nrep} connect the lifted three-body moments to the two-body moments.  These maps are missing from the v2RDM $\mathrm{DQGT_1T_2'}$ consistency conditions. In the next section we will show that degree-$2$  multiplier polynomials are sufficient when the full ${}_{\text{wn}}\Sigma_{d, \leq 2}$ set of constraints are used.

\subsection{Degree-2 multipliers are sufficient}
\label{sec:degree-two-multipliers}
The numerical agreement between the degree-at-most-two and degree-at-most-four multiplier formulations is a consequence of the complete cubic SOS construction, rather than a special property of the Richardson Hamiltonian. We first prove equality of their supremal energy bounds at $d=3$. Appendix~\ref{app:generalized_sufficient_degree2} extends the argument to arbitrary $d\geq2$, for which the full multiplier cutoff is $2d-2$.

For fixed $H$ and $r=N-\eta\mathbb{1}$, define the supremal
weighted SOS energy bound
\begin{align}\label{eq:supremal_energy_def}
E_{d,m}^{*}
:=
\sup_{E\in\mathbb{R},\,S,\,f}
\left\{
E :
\begin{array}{l}
H-E\mathbb{1}=S+W_N(f),\\
S\in\Sigma_d,\quad f\in\mathcal{V}_m,\\
f=f^\dagger,\quad [f,N]=0
\end{array}
\right\},
\end{align}
where $d$ is the SOS generator degree cutoff, $m$ is the
multiplier degree cutoff, and $W_N(f)=fr+rf^\dagger$.
We consider $d\geq2$ and $0\leq m\leq2d-2$.
Since $\mathcal{V}_2\subseteq\mathcal{V}_4$, enlarging the
multiplier space gives
\begin{align}
E_{3,2}^{*}\leq E_{3,4}^{*}.
\end{align}
To establish the reverse inequality, we convert every feasible
certificate with multiplier degree at most four into a certificate
with multiplier degree at most two, keeping the SOS level fixed
and allowing an arbitrarily small decrease in the certified
energy. Our goal is to prove the following:

\begin{proposition}[Degree-two multiplier sufficiency]
\label{prop:degree-two-sufficiency}
For a Hermitian, particle-conserving, at-most-two-body Hamiltonian on finitely
many modes and any particle number $\eta\in\{0,\ldots,R\}$,
\begin{equation}
 E_{3,2}^{*} = E_{3, 4}^{*}
\label{eq:degree-two-equality}
\end{equation}
when the SOS part is generated by the full space $\mathcal{V}_{3}$.
\end{proposition}
We convert any feasible certificate with a degree-at-most-four
multiplier into a certificate with a degree-at-most-two multiplier,
allowing an arbitrarily small decrease in the certified energy.
The quartic multiplier shift is first written as cross terms
between generators in $\mathcal{V}_3$. Completing the squares
and adding an explicit CAR-based SOS complement produces a new
SOS operator and a degree-at-most-two multiplier. The final
assembled SOS has no sextic component.
\begin{proof}
Fix any feasible certificate
\begin{equation}
 H-E\mathbb{1}
 =S+W_N(f_{\leq2})+W_N(f_{(4)}),
 \qquad S\in\Sigma_3,
\label{eq:degree-two-start}
\end{equation}
where $f_{\leq2}$ and $f_{(4)}$ are the Hermitian degree-at-most-two
and quartic parts of Eq.~\eqref{eq:number-multiplier}.

\emph{1. Express the quartic shift as cubic cross terms.}
Define
\begin{align}
 \phi_i&=\sum_{pqs}\bigl(F^{(4)}_{pq;is}\bigr)^*
                  a_s^\dagger a_q a_p,
 \qquad Z_i=a_i r,\qquad Y_i=a_i^\dagger r.
\label{eq:degree-two-generators}
\end{align}
These belong to $\mathcal V_3$: for example,
$Z_i=(1-\eta)a_i+\sum_s a_s^\dagger a_s a_i$ and
$Y_i=\sum_s a_i^\dagger a_s^\dagger a_s-\eta a_i^\dagger$.
The conjugation in $\phi_i$ ensures
$\sum_i\phi_i^\dagger Z_i=f_{(4)}r$. Hence
\begin{equation}
 W_N(f_{(4)})
 =\sum_i\bigl(\phi_i^\dagger Z_i+Z_i^\dagger\phi_i\bigr).
\label{eq:degree-two-cross-terms}
\end{equation}
Both $\phi_i$ and $Z_i$ have charge $\Delta N=-1$, so their cross
terms lie within an allowed Gram block.

\emph{2. Complete the squares and isolate two corrections.}
For any $\lambda>0$, set
$P_i(\lambda)=\lambda^{-1/2}\phi_i+\lambda^{1/2}Z_i$.
The CAR give
\begin{equation}
 Z_i^\dagger Z_i+Y_i^\dagger Y_i
 =r(a_i^\dagger a_i+a_i a_i^\dagger)r=r^2.
\label{eq:degree-two-car-penalty}
\end{equation}
Expanding $P_i^\dagger P_i$ and using this identity yields
\begin{align}
 W_N(f_{(4)})&=Q_\lambda-\lambda Rr^2-\lambda^{-1}K,
 \label{eq:degree-two-square-completion}\\
 Q_\lambda&=\sum_i P_i^\dagger P_i
                 +\lambda\sum_iY_i^\dagger Y_i\in\Sigma_3,
 \qquad K=\sum_i\phi_i^\dagger\phi_i.
 \nonumber
\end{align}
where $R$ is the number of single-particle modes. The term $-\lambda Rr^2$ can be absorbed into a degree-two
multiplier. The remaining negative term $-K/\lambda$ cannot simply
be discarded, even when its coefficient is small.

\emph{3. Replace the negative residual by an SOS minus a constant.}
We construct a finite $C\geq0$ such that
$D=C\mathbb{1}-K\in\Sigma_3$. This requires an SOS decomposition,
not merely the operator inequality $K\preceq C\mathbb{1}$.
For a cubic monomial $B=a_s^\dagger a_q a_p$, successive applications
of $\{a_s,a_s^\dagger\}=\mathbb{1}$, then the analogous identities
for $q$ and $p$, give
\begin{align}
 \mathbb{1}-B^\dagger B
 ={}&(a_p^\dagger)^\dagger a_p^\dagger
 +(a_q^\dagger a_p)^\dagger(a_q^\dagger a_p)\nonumber\\
 &+(a_s a_q a_p)^\dagger(a_s a_q a_p)\in\Sigma_3.
\label{eq:degree-two-monomial-complement}
\end{align}
For a more in-depth description of how this works for general ladder operators $B$ of arbitrary degree see Appendix~\ref{app:generalized_sufficient_degree2}.
Collect the cubic monomials $B_{\alpha}$ occurring in the $\phi_i$ into
$\vec{B}=(B_1,\ldots,B_{n_B})^T$ and write
$\phi_i=\sum_\alpha c_{i\alpha}B_\alpha$ where $c_{i\alpha}$ are the coefficients associated with the monomial in $\phi_{i}$. Then
\begin{equation}
 K=\vec{B}^\dagger G_{K}  \vec{B},\qquad
 \left[G_{K}\right]_{\alpha\beta}=\sum_i c_{i\alpha}^*c_{i\beta},\qquad G_{K}\succeq0.
\end{equation}
Choose $t=\operatorname{Tr}(G_{K})$, so that $tI_{n_B}-G_{K}\succeq0$, and
$C=n_Bt$. Here $I_{n_B}$ is the identity in coefficient space. The exact
identity
\begin{align}
 D:=C\mathbb{1}-K
 ={}&t\sum_{\alpha=1}^{n_B}
       (\mathbb{1}-B_\alpha^\dagger B_\alpha)\nonumber\\
 &+ \vec{B}^\dagger(tI_{n_B}-G_{K}) \vec{B}\in\Sigma_3
\label{eq:degree-two-residual-complement}
\end{align}
proves the claim: the first term uses
Eq.~\eqref{eq:degree-two-monomial-complement}, and the second has a
positive semidefinite Gram matrix. If $f_{(4)}=0$, no conversion is
needed. Otherwise, $C$ is fixed by the original quartic coefficients
and the chosen monomial list, independently of $\lambda$.

\emph{4. Assemble the certificate and compare optimal values.}
Define
\begin{equation}
 \widetilde f=f_{\leq2}-\frac{\lambda R}{2}r,
 \qquad
 \widetilde S=S+Q_\lambda+\frac{D}{\lambda}\in\Sigma_3.
\end{equation}
Substitution into Eq.~\eqref{eq:degree-two-start} gives
\begin{equation}
 H-\left(E-\frac{C}{\lambda}\right)\mathbb{1}
 =\widetilde S+W_N(\widetilde f),
 \qquad \deg\widetilde f\leq2.
\label{eq:degree-two-converted-certificate}
\end{equation}
In coefficient form,
$\widetilde f_0=f_0+\lambda R\eta/2$,
$\widetilde F^{(2)}_{i;j}=F^{(2)}_{i;j}-(\lambda R/2)\delta_{ij}$,
and $\widetilde F^{(4)}=0$.
The converted energy is one feasible degree-two value, not its
optimized value. Since that problem maximizes $E$,
\begin{equation}
 E-\frac{C}{\lambda}\leq E_{3,2}^*\leq E_{3,4}^*
 \qquad\text{for every }\lambda>0.
\label{eq:degree-two-sandwich}
\end{equation}
Keeping the original certificate, and therefore $E$ and $C$, fixed,
we may take $\lambda\to\infty$ to obtain $E_{3,2}^{*}\geq E$.
Because $E$ was any feasible degree-four value, taking the supremum
over those values gives $E_{3,2}^{*}\geq E_{3,4}^{*}$ and proves equality.
\end{proof}
The equality $E_{3,2}^{*}=E_{3,4}^{*}$ concerns supremal
energy bounds; the construction does not establish a
finite-coefficient conversion at the original energy $E$. In particular, $\widetilde f=f_{\leq2}-\lambda Rr/2$ generally grows with $\lambda$, even though the energy loss $C/\lambda$ tends to zero. Within a finite-precision SDP calculation, the relevant comparison
is at a prescribed energy tolerance $\epsilon>0$. For any fixed
feasible degree-four-multiplier certificate with energy $E$,
choosing a finite $\lambda>C/\epsilon$ produces a feasible
degree-two-multiplier certificate with finite coefficients and
energy $\widetilde E_\lambda=E-C/\lambda>E-\epsilon$.
Thus, in exact arithmetic, removing the quartic multipliers
 does not introduce unavoidable loss at any prescribed energy tolerance.

Equation~\eqref{eq:degree-two-converted-certificate} implies
$\widetilde S=H-(E-C/\lambda)\mathbb{1}-W_N(\widetilde f)$,
which has degree at most four. If $\widetilde G\succeq0$ is the Gram matrix satisfying $\widetilde S=\Phi_3(\widetilde G)$ then the total sextic coefficient
vanishes exactly, $ A_6\operatorname{vec}(\widetilde G)=0$, by construction. In other words, the new SOS retains the full nullspace
condition, while eliminating the quartic multiplier variables.
It uses no special properties of the Hamiltonian beyond Hermiticity, particle conservation, and ladder degree at most four. However, it requires that the full SOS generator algebra be available. The proof for the collapse to degree-2 multiplier polynomials can be generalized to ${}_{\text{wn}}\Sigma_{d, \leq 2}$ using the same techniques as above. Details are provided in Appendix~\ref{app:generalized_sufficient_degree2}.

\section{Discussion}
We have presented a protocol and software for constructing all witnesses to verify that a given $2$-RDM is $n$-representable based on the weighted SOS of Ref.~\cite{pironio2010convergent, helton2004positivstellensatz}. This work instantiates a Gram matrix cancellation program anticipated by Refs.~\cite{PhysRevLett.108.263002, PhysRevLett.130.153001, yt4s-dlwr}, making it rigorous and complete in the operator algebra formalism of wSOS. By detailing the conversion to the primal (v2RDM) program we reveal the additional condition that RDMs must satisfy to be approximately $n$-representable.

The purpose of this work was to provide an implementable methodology based on wSOS to construct 2-body witnesses despite the expected exponential scaling of a full enumeration. With such a tool further exploration of high-order cancellation conditions and wSOS for quantum and classical algorithms is possible~\cite{rubin2026near, pb2g-j9cw}. For example, one can imagine using the entire ${}_{\operatorname{wn}}\Sigma_{\eta, \leq 2}$ but variationally selecting some function of the operators like in Ref~\cite{PhysRevX.14.021008}. That would require a way to evaluate the cancellation conditions without explicit construction of the linear maps over objects in ${}_{\operatorname{wn}}\Sigma_{\eta, \leq 2}$.   

\subsection{AI Statement}
This work builds on the SOS representations of Ref.~\cite{rubin2026near}. The key components of the constructive solution are the work of humans, but some key proofs come from LLMs. Specifically, the reduction of degree-$4$ multiplier polynomials to degree-$2$ multiplier polynomials and the generalized form in Appendix~\ref{app:generalized_sufficient_degree2} along with the proof for the sufficiency of ${}_{\operatorname{wn}}\Sigma_{\eta, \leq 2}$ in Section~\ref{sec:complete_nrep_phys_witness} were derived by OpenAI's Astra-6 pro model. Early discussions with Gemini 3.1 pro indicated that weighted SOS could be used to construct a complete solution. Gemini 4 and Astra aided in writing the manuscript. All authors assume full responsibility for the correctness of the results.

\bibliography{reference}

\appendix

\section{Standard RDM positivity conditions as examples of the adjoint construction}
\label{app:standard-rdm-adjoint-examples}

The abstract primal constraint
\begin{align}
    M(x,u)
    =
    \operatorname{vec}^{-1}
    \left(
        A_{\leq4}^{T}x
        +
        A_{>4}^{T}u
    \right)
    \succeq0
    \label{eq:appendix-general-lifted-moment}
\end{align}
contains the familiar $D$, $Q$, $G$, and $T_2$ positivity conditions
as special cases.  The purpose of this appendix is to make this
connection explicit.

Let $\mathcal L_x$ denote the linear functional specified by the
canonical low-body moments
\begin{align}
    x_0
    &=
    \mathcal L_x(\mathbb 1),
    \\
    {}^{1}D^{i}_{j}
    &=
    \mathcal L_x
    \left(
        a_i^\dagger a_j
    \right),
    \\
    {}^{2}D^{ij}_{kl}
    &=
    \mathcal L_x
    \left(
        a_i^\dagger a_j^\dagger a_l a_k
    \right).
    \label{eq:appendix-canonical-low-moments}
\end{align}
The normalization condition fixes $x_0=1$.

For a chosen generator family
\begin{align}
    \vec{o}_s
    =
    \begin{pmatrix}
        o^{(s)}_1 & \cdots & o^{(s)}_{L_s}
    \end{pmatrix}^{T},
\end{align}
let $K_s\succeq0$ denote the corresponding \emph{dual} Gram matrix.
The SOS operator is
\begin{align}
   W_s(K_s)
    =
    \vec{o}_s^{\dagger}
    K_s
    \vec{o}_s.
\end{align}
Normal ordering defines coefficient maps
\begin{align}
    A_{s,\leq4}
    \operatorname{vec}(K_s)
    &=
    \text{degree-$\leq4$ coefficients of }  W_s(K_s),
    \\
    A_{s,>4}
    \operatorname{vec}(K_s)
    &=
    \text{degree-$>4$ coefficients of } W_s(K_s).
\end{align}
The adjoint maps assemble the primal moments into a matrix on the same
generator-index space as $K_s$:
\begin{align}
    M_s(x,u)
    =
    \operatorname{vec}^{-1}
    \left(
        A_{s,\leq4}^{T}x
        +
      A_{s,>4}^{T}u
    \right).
    \label{eq:appendix-family-moment-map}
\end{align}
Its defining relation is
\begin{align}
    \left\langle M_s(x,u),K_s\right\rangle
    =
    x^T  A_{s,\leq4}\operatorname{vec}(K_s)
    +
    u^T  A_{s,>4}\operatorname{vec}(K_s).
    \label{eq:appendix-adjoint-pairing}
\end{align}
Equivalently, its entries may be viewed as the moments of the generator
products,
\begin{align}
    [M_s(x,u)]_{\alpha\beta}
    =
    \mathcal L_{x,u}
    \left(
        o^{(s)\dagger}_{\alpha}
        o^{(s)}_{\beta}
    \right),
    \label{eq:appendix-moment-entry}
\end{align}
where $u$ supplies auxiliary values for any canonical operators of
degree greater than four.

The positivity condition
\begin{align}
    M_s(x,u)\succeq0
\end{align}
is equivalent to requiring
\begin{align}
    \mathcal L_{x,u}
    \left(
        R^\dagger R
    \right)
    \geq0
\end{align}
for every $R$ in the span of $\boldsymbol{o}_s$.

\subsection{The \texorpdfstring{$D$}{D} condition}

Choose the pair-annihilation generators
\begin{align}
    d_{ij}=a_j a_i,
\end{align}
where the composite pair index may be restricted to $i<j$.  Their
products are
\begin{align}
    d_{ij}^\dagger d_{kl}
    =
    a_i^\dagger a_j^\dagger a_l a_k,
\end{align}
which are already canonical degree-four operators.  Consequently,
\begin{align}
    A_{D,>4}=0.
\end{align}
The adjoint-assembled moment matrix has entries
\begin{align}
    [M_D(x)]_{ij,kl}
    &=
    \mathcal L_x
    \left(
        d_{ij}^\dagger d_{kl}
    \right)
    \nonumber\\
    &=
    \mathcal L_x
    \left(
        a_i^\dagger a_j^\dagger a_l a_k
    \right)
    \nonumber\\
    &=
    {}^{2}D^{ij}_{kl}.
\end{align}
Therefore,
\begin{align}
    M_D(x)
    =
    \operatorname{vec}^{-1}
    \left(
         A_{D,\leq4}^{T}x
    \right)
    =
    {}^{2}D,
\end{align}
and the general primal constraint reduces to the usual $D$ condition
\begin{align}
    {}^{2}D\succeq0.
    \label{eq:appendix-D-condition}
\end{align}

\subsection{The \texorpdfstring{$Q$}{Q} condition}

Choose the pair-creation generators
\begin{align}
    q_{ij}=a_j^\dagger a_i^\dagger,
\end{align}
so that
\begin{align}
    q_{ij}^\dagger q_{kl}
    =
    a_i a_j a_l^\dagger a_k^\dagger.
\end{align}
Normal ordering gives
\begin{align}
    a_i a_j a_l^\dagger a_k^\dagger
    ={}&
    \left(
        \delta_{ik}\delta_{jl}
        -
        \delta_{il}\delta_{jk}
    \right)\mathbb 1
    \nonumber\\
    &-
    \delta_{jl}a_k^\dagger a_i
    +
    \delta_{il}a_k^\dagger a_j
    +
    \delta_{jk}a_l^\dagger a_i
    -
    \delta_{ik}a_l^\dagger a_j
    \nonumber\\
    &+
    a_k^\dagger a_l^\dagger a_j a_i.
    \label{eq:appendix-Q-normal-order}
\end{align}
No operator of degree greater than four is generated, and hence
\begin{align}
    A_{Q,>4}=0.
\end{align}
Applying $\mathcal L_x$ to
Eq.~\eqref{eq:appendix-Q-normal-order} gives
\begin{align}
    [M_Q(x)]_{ij,kl}
    ={}&
    \left(
        \delta_{ik}\delta_{jl}
        -
        \delta_{il}\delta_{jk}
    \right)x_0
    \nonumber\\
    &-
    \delta_{jl}{}^{1}D^{k}{}_{i}
    +
    \delta_{il}{}^{1}D^{k}{}_{j}
    +
    \delta_{jk}{}^{1}D^{l}{}_{i}
    -
    \delta_{ik}{}^{1}D^{l}{}_{j}
    \nonumber\\
    &+
    {}^{2}D^{kl}_{ij}.
    \label{eq:appendix-Q-from-D}
\end{align}
This is the conventional two-hole matrix,
\begin{align}
    M_Q(x)
    =
    \operatorname{vec}^{-1}
    \left(
         A_{Q,\leq4}^{T}x
    \right)
    =
    {}^{2}Q.
\end{align}
Thus the adjoint-map constraint is exactly
\begin{align}
    {}^{2}Q\succeq0.
    \label{eq:appendix-Q-condition}
\end{align}

\subsection{The particle--hole \texorpdfstring{$G$}{G} condition}

Choose the particle--hole generators
\begin{align}
    g_{ij}=a_j^\dagger a_i.
\end{align}
To avoid a notation collision, $K_G$ denotes the dual SOS Gram matrix,
while ${}^{2}G$ denotes the primal particle--hole metric matrix.

The generator products satisfy
\begin{align}
    g_{ij}^\dagger g_{kl}
    &=
    a_i^\dagger a_j a_l^\dagger a_k
    \nonumber\\
    &=
    \delta_{jl}a_i^\dagger a_k
    -
    a_i^\dagger a_l^\dagger a_j a_k.
    \label{eq:appendix-G-normal-order}
\end{align}
Again no operator of degree greater than four appears, so
\begin{align}
    A_{G,>4}=0.
\end{align}
The moment matrix is
\begin{align}
    [M_G(x)]_{ij,kl}
    &=
    \mathcal L_x
    \left(
        g_{ij}^\dagger g_{kl}
    \right)
    \nonumber\\
    &=
    \delta_{jl}{}^{1}D^{i}{}_{k}
    -
    {}^{2}D^{il}_{kj}.
    \label{eq:appendix-G-from-D}
\end{align}
This is the usual particle--hole matrix:
\begin{align}
    M_G(x)
    =
    \operatorname{vec}^{-1}
    \left(
         A_{G,\leq4}^{T}x
    \right)
    =
    {}^{2}G.
\end{align}
Therefore the general primal constraint becomes
\begin{align}
    {}^{2}G\succeq0.
    \label{eq:appendix-G-condition}
\end{align}

\subsection{The \texorpdfstring{$T_2$}{T2} condition}

The $T_2$ condition is the first example in which the generator
products contain operators of degree greater than four.  Let
\begin{align}
    X_I
    =
    a_i^\dagger a_j^\dagger a_k,
    \qquad
    I=(i,j,k).
\end{align}
Particle conservation gives two dual Gram blocks,
\begin{align}
    K_+\succeq0,
    \qquad
    K_-\succeq0,
\end{align}
which weight the two operator orientations
\begin{align}
    X_I^\dagger X_J,
    \qquad
    X_JX_I^\dagger.
\end{align}
After aligning the composite indices, their normal-ordered forms may
be written as
\begin{align}
    X_I^\dagger X_J
    &=
    E^{(6)}_{IJ}
    +
    L^+_{IJ},
    \\
    X_JX_I^\dagger
    &=
    -E^{(6)}_{IJ}
    +
    L^-_{IJ},
    \label{eq:appendix-T2-two-orientations}
\end{align}
where $E^{(6)}_{IJ}$ is the common degree-six operator and
$L^\pm_{IJ}$ contain only one- and two-body operators.

A sufficient way to cancel the sextic contributions is to
impose the stronger coefficient-matching condition
$K_+=\widetilde K_-$. This defines the conventional
$T_2$ subcone; it is not the full canonical sextic nullspace:
\begin{align}
    A_{\text{tie}}(K_+,\widetilde K_-)
    =
    K_+-\widetilde K_-,
    \label{eq:appendix-T2-A6}
\end{align}
where $\widetilde K_-$ denotes the index-aligned transpose of $K_-$.
The nullspace condition is
\begin{align}\label{eq:app-T2-k-equals}
    K_+-\widetilde K_-=0.
\end{align}

Define the low-body moment matrices
\begin{align}
    [M_+(x)]_{IJ}
    &=
    \mathcal L_x
    \left(
        L^+_{IJ}
    \right),
    \\
    [M_-(x)]_{IJ}
    &=
    \mathcal L_x
    \left(
        L^-_{IJ}
    \right).
\end{align}
Before eliminating the multiplier for the Gram equality, the general
adjoint construction gives
\begin{align}
    M_{T_2}^{\mathrm{lift}}(x,U)
    =
    \begin{pmatrix}
        M_+(x)-U & 0\\
        0 & M_-(x)+U
    \end{pmatrix}
    \succeq0,
    \label{eq:appendix-T2-lifted}
\end{align}
where $U$ is an unrestricted Hermitian Lagrange multiplier
for the Gram equality in Eq.~\eqref{eq:app-T2-k-equals}; it is not restricted
to the range of the canonical sextic adjoint map.

Equation~\eqref{eq:appendix-T2-lifted} is equivalent to
\begin{align}
    M_+(x)-U&\succeq0,
    \\
    M_-(x)+U&\succeq0.
\end{align}
Adding these inequalities gives
\begin{align}
    M_+(x)+M_-(x)\succeq0.
    \label{eq:appendix-T2-sum}
\end{align}
Conversely, if Eq.~\eqref{eq:appendix-T2-sum} holds, choosing
\begin{align}
    U=M_+(x)
\end{align}
makes the first lifted block equal to zero and the second equal to
$M_+(x)+M_-(x)$.  Therefore,
\begin{align}
    &\exists\,U:
    \begin{cases}
        M_+(x)-U\succeq0,\\
        M_-(x)+U\succeq0
    \end{cases}
    \nonumber\\
    &\hspace{2cm}\Longleftrightarrow\quad
    M_+(x)+M_-(x)\succeq0.
    \label{eq:appendix-T2-elimination-equivalence}
\end{align}

Using Eq.~\eqref{eq:appendix-T2-two-orientations}, the entries of the
summed matrix are
\begin{align}
    [M_+(x)+M_-(x)]_{IJ}
    &=
    \mathcal L_x
    \left(
        L^+_{IJ}+L^-_{IJ}
    \right)
    \nonumber\\
    &=
    \mathcal L_x
    \left(
        X_I^\dagger X_J
        +
        X_JX_I^\dagger
    \right)
    \nonumber\\
    &=
    \mathcal L_x
    \left(
        \left\{
            X_I^\dagger,X_J
        \right\}
    \right).
\end{align}
Thus
\begin{align}
    T_2(x)
    :=
    M_+(x)+M_-(x),
\end{align}
and the simplified primal condition is the familiar LMI
\begin{align}
    T_2(x)\succeq0.
    \label{eq:appendix-T2-condition}
\end{align}
The auxiliary degree-six variable has not been set to zero.  It has
been eliminated exactly because the truncation map has the special
difference structure
\begin{align}
    A_{\text{tie}}=(I,-I).
\end{align}

\subsection{Relation to the conventional v2RDM formulation}

If the four generator families are kept as separate Gram blocks, the
unsimplified lifted primal matrix is
\begin{align}
    M_{\mathrm{DQGT}_2}^{\mathrm{lift}}(x,U)
    ={}&
    {}^{2}D(x)
    \oplus
    {}^{2}Q(x)
    \oplus
    {}^{2}G(x)
    \nonumber\\
    &\oplus
    \left[M_+(x)-U\right]
    \oplus
    \left[M_-(x)+U\right].
    \label{eq:appendix-DQGT2-lifted-direct-sum}
\end{align}
The condition
\begin{align}
    M_{\mathrm{DQGT}_2}^{\mathrm{lift}}(x,U)\succeq0
\end{align}
is therefore equivalent, after eliminating $U$, to
\begin{align}
    {}^{2}D&\succeq0,
    \\
    {}^{2}Q&\succeq0,
    \\
    {}^{2}G&\succeq0,
    \\
    T_2&\succeq0.
    \label{eq:appendix-conventional-DQGT2}
\end{align}

In the canonical-moment parameterization used above, the scalar,
1-RDM, and 2-RDM moments are stored once in $x$, and the matrices
${}^{2}D$, ${}^{2}Q$, ${}^{2}G$, and $T_2$ are assembled by the
adjoint coefficient maps.  In the conventional v2RDM program, these
metric matrices are often introduced as separate PSD blocks, while
affine mapping conditions enforce Eqs.~\eqref{eq:appendix-Q-from-D}
and~\eqref{eq:appendix-G-from-D}, together with the corresponding
normal-ordering and contraction identities.  These are two equivalent
coordinate descriptions of the same positivity constraints.

The $D$, $Q$, and $G$ conditions are the simplest adjoint-map examples
because their generator products never exceed degree four:
\begin{align}
    A_{D,>4}
    =
     A_{Q,>4}
    =
    A_{G,>4}
    =
    0.
\end{align}
The $T_2$ condition is the simplest nontrivial nullspace example:
$A_{\text{tie}}\neq0$, but the corresponding auxiliary multiplier can be
eliminated analytically.  For general $\Sigma_{d,\leq 2}$ conditions, the
high-body maps couple many Gram blocks through permutations and
contractions, and the lifted variables in
Eq.~\eqref{eq:appendix-general-lifted-moment} need not reduce to a
single named RDM metric matrix.

\subsection{(2,4)-conditions}

\paragraph{General quartic truncation equations.}
For a quartic pattern
$\sigma=(\sigma_1,\ldots,\sigma_4)\in\{x,o\}^4$, define
\begin{equation}
    A_\sigma(I)
    =
     a_{i_1}^{\sigma_1}
     a_{i_2}^{\sigma_2}
     a_{i_3}^{\sigma_3}
     a_{i_4}^{\sigma_4},
    \qquad
    a_i^x= a_i^\dagger,
    \quad
     a_i^o= a_i,
\end{equation}
where $I=(i_1,i_2,i_3,i_4)$.  Let
\begin{equation}
    \mathcal{S}_\eta
    =
    \{\sigma\in\{x,o\}^4:\Delta N(\sigma)=\eta\}.
\end{equation}
The most general particle-conserving quartic Gram operator is
\begin{equation}
     W(G)
    =
    \sum_\eta
    \sum_{\sigma,\tau\in\mathcal{S}_\eta}
    \sum_{I,J}
    G_{\sigma,\tau}(I;J)
     A_\sigma(I)
     A_\tau(J)^\dagger,
    \qquad
    G\succeq0.
    \label{eq:general-quartic-gram-two-four}
\end{equation}
For a canonical degree-$m$ operator $E_m(y)$, let
$\mathcal K_m(\sigma,\tau)$ denote the normal-ordering branches of
$ A_\sigma(I) A_\tau(J)^\dagger$ that leave $m$
ladder operators.  If $\kappa\in\mathcal K_m(\sigma,\tau)$ has CAR
coefficient $\Gamma_{\sigma,\tau,\kappa}^{(m)}$ and index-placement
map
\begin{equation}
    R_{\sigma,\tau,\kappa}^{(m)}(y, u):(y, u)\longmapsto(I;J),
\end{equation}
then the coefficient of $ E_m(y)$ is
\begin{equation}
    q_m(y)
    =
    \sum_\eta
    \sum_{\sigma,\tau\in\mathcal{S}_\eta}
    \sum_{\kappa\in\mathcal K_m(\sigma,\tau)}
    \Gamma_{\sigma,\tau,\kappa}^{(m)}
    \sum_{ u}
    G_{\sigma,\tau}
    \!\left(
        R_{\sigma,\tau,\kappa}^{(m)}(y, u)
    \right).
    \label{eq:general-qm-two-four}
\end{equation}
For a no-contraction degree-eight branch,
$R_{\sigma,\tau}^{(8)}=(P_{\sigma,\tau}^{(8)})^{-1}$ and the sum over
$ u$ is absent.  The quartic SOS operator is evaluable from at
most two-particle information precisely when
\begin{equation}
    q_8(y)=0,
    \qquad
    q_6(y)=0,
    \qquad
    \text{for every canonical }y.
    \label{eq:full-A8-A6-condition}
\end{equation}
These equations are the symbolic form of
$A_8\operatorname{vec}(G)=0$ and
$A_6\operatorname{vec}(G)=0$.

\paragraph{Mazziotti's factorized diagonal ansatz.}
Following Ref.~\cite{PhysRevLett.108.263002}, let $v_1=b$, $v_2=c$, $v_3=d$, and $v_4=e$, and define
\begin{equation}
     C_\sigma
    =
    \sum_I
    c_\sigma(I) A_\sigma(I),
    \qquad
    c_\sigma(I)
    =
    \prod_{a=1}^4 v_{a,i_a}^{\sigma_a},
    \label{eq:C-sigma-factorized}
\end{equation}
where
\begin{equation}
    v_{a,i}^{x}=v_{a,i}^*,
    \qquad
    v_{a,i}^{o}=v_{a,i}.
\end{equation}
Write
\begin{equation}
     S_\sigma
    :=
     C_\sigma C_\sigma^\dagger
    \succeq0,
    \qquad
    K_\sigma(I;J)
    :=
    c_\sigma(I)c_\sigma(J)^*.
\end{equation}
For
\begin{equation}
    y=(\alpha;\beta),
    \qquad
    \alpha=(\alpha_1,\ldots,\alpha_4),
    \quad
    \beta=(\beta_1,\ldots,\beta_4),
\end{equation}
define $P_\sigma$ slotwise by
\begin{equation}
    (\alpha_a,\beta_a)
    =
    \begin{cases}
        (i_a,j_a), & \sigma_a=x,\\
        (j_a,i_a), & \sigma_a=o.
    \end{cases}
    \label{eq:P-sigma-slotwise}
\end{equation}
Thus $P_\sigma^{-1}(y)$ supplies the raw $(I;J)$ arguments that
contribute to the common canonical operator labelled by $y$.  In each
$x$ slot the local coefficient is
$v_{a,\alpha_a}^*v_{a,\beta_a}$, while in each $o$ slot it is
$v_{a,\beta_a}v_{a,\alpha_a}^*$, which is the same scalar.  Therefore
\begin{equation}
    K_\sigma\!\left(P_\sigma^{-1}(y)\right)
    =
    Q_8(y)
    :=
    \prod_{a=1}^4
    v_{a,\alpha_a}^*v_{a,\beta_a},
    \label{eq:common-Q8-unified}
\end{equation}
independently of $\sigma$.  If
$z_{\boldsymbol\alpha}=\prod_{a=1}^4v_{a,\alpha_a}$, then
\begin{equation}
    Q_8(\boldsymbol\alpha;\boldsymbol\beta)
    =
    z_{\boldsymbol\alpha}^*z_{\boldsymbol\beta},
    \qquad
    Q_8(\boldsymbol\beta;\boldsymbol\alpha)
    =
    Q_8(\boldsymbol\alpha;\boldsymbol\beta)^*,
    \qquad
    Q_8\succeq0.
    \label{eq:Q8-Hermitian}
\end{equation}
Hence the coefficient tensor is Hermitian in the two composite indices;
it is ordinarily symmetric only when the vectors may be chosen real.
The indices in $y$ are initially labelled by the slots $b,c,d,e$.
Any additional antisymmetrization required to sort the creation and
annihilation tuples is common to every pattern and therefore does not
alter the cancellations below.

A contraction of annihilator slot $r$ with creator slot $s$ identifies
$\beta_r=\alpha_s=u$ and produces the common degree-six tensor
\begin{equation}
    Q_6^{rs}
    (\boldsymbol\alpha_{ s};
     \boldsymbol\beta_{r})
    :=
    \sum_u
    Q_8
    (\alpha_1,\ldots,\alpha_s=u,\ldots,\alpha_4;
     \beta_1,\ldots,\beta_r=u,\ldots,\beta_4)
    \label{eq:common-Q6-unified}
\end{equation}
\begin{equation}
    \hspace{2.6cm}
    =
    \langle v_s,v_r\rangle
    \prod_{a\neq s}v_{a,\alpha_a}^*
    \prod_{b\neq r}v_{b,\beta_b}.
\end{equation}
Thus factorization makes both the degree-eight tensor and every fixed
degree-six contraction tensor independent of the pattern; only the CAR
coefficients and the external square weights depend on $\sigma$.

\paragraph{Derivation of the weights.}
Consider the six-pattern support of the first row of Table~I of Ref.~\cite{PhysRevLett.108.263002},
\begin{equation}
\begin{aligned}
     W(w)
    ={}&
    w_p S_{xxxx}
    +w_4 S_{xxxo}
    +w_3  S_{xxox}
    +w_2  S_{xoxx}
    +w_1 S_{oxxx}
    +w_h S_{oooo},
\end{aligned}
    \label{eq:weighted-seed-unified}
\end{equation}
where $w_p,w_h,w_a\geq0$.  Equivalently, let
$\sigma^{(a)}$ denote the pattern with a single $o$ in slot $a$:
\begin{equation}
    \sigma^{(1)}=oxxx,
    \qquad
    \sigma^{(2)}=xoxx,
    \qquad
    \sigma^{(3)}=xxox,
    \qquad
    \sigma^{(4)}=xxxo.
\end{equation}
For the same-slot contraction $(r,s)=(a,a)$, direct CAR normal ordering
gives coefficient $+1$ for $ S_{\sigma^{(a)}}$, coefficient
$-1$ for $ S_{oooo}$, and zero for the other four patterns.
Consequently,
\begin{equation}
    q_6^{aa}(y)
    =
    (w_a-w_h)Q_6^{aa}(y).
\end{equation}
Cancellation for arbitrary $b,c,d,e$ therefore requires
\begin{equation}
    w_1=w_2=w_3=w_4=w_h.
    \label{eq:A6-weight-equalities}
\end{equation}
For $r\neq s$, the corresponding CAR coefficient has the form
\begin{equation}
    q_6^{rs}(y)
    =
    \varepsilon_{rs}
    \bigl(w_{\min\{r,s\}}-w_h\bigr)
    Q_6^{rs}(y),
    \qquad
    \varepsilon_{rs}\in\{+1,-1\},
\end{equation}
so the remaining contraction channels impose no additional relations.

The no-contraction sign of
$ S_\sigma$ is $(-1)^{n_o(\sigma)}$.  Hence
\begin{equation}
    q_8(y)
    =
    \left(
        w_p-\sum_{a=1}^4w_a+w_h
    \right)Q_8(y).
\end{equation}
Using Eq.~\eqref{eq:A6-weight-equalities}, the degree-eight equation is
\begin{equation}
    (w_p-3w_h)Q_8(y)=0,
\end{equation}
which implies $w_p=3w_h$.  In the ordering
\begin{equation}
    w
    =
    (w_{xxxx},w_{xxxo},w_{xxox},w_{xoxx},w_{oxxx},w_{oooo})^T,
\end{equation}
the reduced factorized truncation equations are therefore
\begin{equation}
    \begin{pmatrix}
         1&-1&-1&-1&-1& 1\\
         0& 1& 0& 0& 0&-1\\
         0& 0& 1& 0& 0&-1\\
         0& 0& 0& 1& 0&-1\\
         0& 0& 0& 0& 1&-1
    \end{pmatrix}
    w
    =0.
    \label{eq:factorized-truncation-matrix-unified}
\end{equation}
Its nullspace is one dimensional:
\begin{equation}
    \ker A_{\mathrm{fact}}
    =
    \operatorname{span}
    \left\{
        (3,1,1,1,1,1)^T
    \right\}.
    \label{eq:factorized-weight-ray-unified}
\end{equation}
After fixing the irrelevant overall scale, the first condition is
\begin{equation}
     W_1
    =
    3 S_{xxxx}
    + S_{xxxo}
    + S_{xxox}
    + S_{xoxx}
    + S_{oxxx}
    + S_{oooo}.
    \label{eq:first-table-I-condition-unified}
\end{equation}
The factor $3$ is therefore fixed jointly by the degree-six equations,
which equate the five unit-weight blocks, and the degree-eight equation,
which then requires $3-4+1=0$.

\paragraph{Generation of all eight Table-I conditions.}
For $F\subseteq\{1,2,3,4\}$, let $\phi_F$ switch
$x\leftrightarrow o$ in every slot belonging to $F$.  Applying the same
slot flip to all six terms permutes the contraction channels and changes
their CAR signs by an invertible signed relabeling; it therefore maps a
solution of the $A_8$ and $A_6$ equations to another solution with the
same weights.  The eight rows of Table~I of Ref.~\cite{PhysRevLett.108.263002} are generated by
\begin{equation}
    F_1=\varnothing,
    \quad
    F_2=\{4\},
    \quad
    F_3=\{3\},
    \quad
    F_4=\{2\},
    \quad
    F_5=\{1\},
    \quad
    F_6=\{3,4\},
    \quad
    F_7=\{2,3\},
    \quad
    F_8=\{2,4\}.
\end{equation}
Equivalently,
\begin{align}
     W_1={}&
    3 S_{xxxx}
    + S_{xxxo}+ S_{xxox}
    + S_{xoxx}+ S_{oxxx}+ S_{oooo},
    \\
     W_2={}&
    3 S_{xxxo}
    + S_{xxxx}+ S_{xxoo}
    + S_{xoxo}+ S_{oxxo}+ S_{ooox},
    \\
      W_3={}&
    3 S_{xxox}
    + S_{xxoo}+ S_{xxxx}
    + S_{xoox}+ S_{oxox}+ S_{ooxo},
    \\
     W_4={}&
    3 S_{xoxx}
    + S_{xoxo}+ S_{xoox}
    + S_{xxxx}+ S_{ooxx}+ S_{oxoo},
    \\
      W_5={}&
    3 S_{oxxx}
    + S_{oxxo}+ S_{oxox}
    + S_{ooxx}+ S_{xxxx}+ S_{xooo},
    \\
      W_6={}&
    3 S_{xxoo}
    + S_{xxox}+ S_{xxxo}
    + S_{xooo}+ S_{oxoo}+ S_{ooxx},
    \\
      W_7={}&
    3 S_{xoox}
    + S_{xooo}+ S_{xoxx}
    + S_{xxox}+ S_{ooox}+ S_{oxxo},
    \\
      W_8={}&
    3 S_{xoxo}
    + S_{xoxx}+ S_{xooo}
    + S_{xxxo}+ S_{ooxo}+ S_{oxox}.
    \label{eq:all-eight-table-I-unified}
\end{align}
For every $r=1,\ldots,8$,
\begin{equation}
     W_r\succeq0,
    \qquad
    \Pi_8( W_r)=0,
    \qquad
    \Pi_6( W_r)=0.
\end{equation}
Thus each $ W_r$ is an at-most-two-body positive operator and
provides the corresponding $(2,4)$ condition.  Switching all four slots
produces its particle--hole dual.

\paragraph{Relation to the full tensor nullspace.}
The preceding derivation uses the restrictive conditions
\begin{equation}
    G_{\sigma,\tau}=0
    \quad(\sigma\neq\tau),
    \qquad
    G_{\sigma,\sigma}
    \text{ rank one within each pattern block},
\end{equation}
together with the shared-tensor identities
\begin{equation}
    G_{\sigma,\sigma}
    \!\left(P_\sigma^{-1}(y)\right)
    =
    w_\sigma Q_8(y)
\end{equation}
and their degree-six contracted analogues.  In the unrestricted Gram
construction, neither equality between the pulled-back tensors nor
pattern diagonality is imposed.  Instead one retains all blocks
$G_{\sigma,\tau}$ satisfying
$\Delta N(\sigma)=\Delta N(\tau)$ and solves the tensor-valued equations
\begin{align}
    0=q_8(y)
    &={}
    \sum_\eta
    \sum_{\sigma,\tau\in\mathcal{S}_\eta}
    \Gamma_{\sigma,\tau}^{(8)}
    G_{\sigma,\tau}
    \!\left(
        \left(P_{\sigma,\tau}^{(8)}\right)^{-1}(y)
    \right),
    \\
    0=q_6(y)
    &={}
    \sum_\eta
    \sum_{\sigma,\tau\in\mathcal{S}_\eta}
    \sum_{\kappa\in\mathcal K_6(\sigma,\tau)}
    \Gamma_{\sigma,\tau,\kappa}^{(6)}
    \sum_{u}
    G_{\sigma,\tau}
    \!\left(
        R_{\sigma,\tau,\kappa}^{(6)}(y,u)
    \right).
    \label{eq:unrestricted-A8-A6-unified}
\end{align}
The complete homogeneous quartic-generator SOS cone that reduces to at most
two-body form is therefore
\begin{equation}
    \mathcal K_{4\rightarrow2}
    =
    \left\{
        G\succeq0:
        q_8(y)=0,\;
        q_6(y)=0
        \text{ for all canonical }y
    \right\}.
    \label{eq:full-quartic-nullspace-cone}
\end{equation}
Mazziotti's Table-I conditions are explicit diagonal, factorized
families inside this larger spectrahedral cone.  They demonstrate one
mechanism for solving the tensor nullspace, but do not exhaust the
solutions allowed by Eqs.~\eqref{eq:general-qm-two-four} and
\eqref{eq:unrestricted-A8-A6-unified}; Ref.~\cite{PhysRevLett.108.263002}
likewise describes the displayed conditions as representative rather than
exhaustive.  A degree-at-most-four generator basis additionally includes
lower-degree and cross-degree Gram blocks, treated by the same truncation
formula in Eq.~\eqref{eq:general-qm-two-four}.

\section{Contraction condition nullspace construction example }\label{app:contraction_condition_construction}
As an example of two cases, consider
\begin{equation}
    \sigma=xxxo,
    \qquad
    \tau=xxox,
\end{equation}
which satisfy
\begin{equation}
    \Delta N(\sigma)
    =
    \Delta N(\tau)
    =
    2.
\end{equation}
Consequently, the cross-pattern Gram block
$G_{xxxo,xxox}$ is particle conserving.

Let
\begin{equation}
    I=(j,k,l,m),
    \qquad
    J=(p,q,r,t).
\end{equation}
The corresponding bare generator words are
\begin{equation}
     A_{xxxo}(I)
    =
     a_j^\dagger
     a_k^\dagger
     a_l^\dagger
     a_m
\end{equation}
and
\begin{equation}
     A_{xxox}(J)
    =
     a_p^\dagger
     a_q^\dagger
     a_r
     a_t^\dagger.
\end{equation}
Taking the Hermitian adjoint reverses the operator order and exchanges
creation and annihilation operators:
\begin{equation}
     A_{xxox}(J)^\dagger
    =
     a_t
     a_r^\dagger
     a_q
     a_p.
\end{equation}
Therefore, the raw product associated with the Gram element
$G_{xxxo,xxox}(I;J)$ is
\begin{equation}
\begin{aligned}
     A_{xxxo}(I)
     A_{xxox}(J)^\dagger
    =
     a_j^\dagger
     a_k^\dagger
     a_l^\dagger
     a_m
     a_t
     a_r^\dagger
     a_q
     a_p.
    \label{eq:xxxo-xxox-raw-product}
\end{aligned}
\end{equation}

The only creation operator that is initially to the right of
annihilation operators is $ a_r^\dagger$.  We first evaluate
the local product
\begin{equation}
     a_m
     a_t
     a_r^\dagger.
\end{equation}
Using the canonical anticommutation relation
\begin{equation}
     a_i a_j^\dagger
    =
    \delta_{ij}
    -
    a_j^\dagger a_i,
\end{equation}
we obtain
\begin{align}
     a_m
     a_t
     a_r^\dagger
    &=
     a_m
    \left(
        \delta_{tr}
        -
        a_r^\dagger a_t
    \right)
    \nonumber\\
    &=
    \delta_{tr} a_m
    -
     a_m
     a_r^\dagger
     a_t
    \nonumber\\
    &=
    \delta_{tr} a_m
    -
    \left(
        \delta_{mr}
        -
        a_r^\dagger a_m
    \right)
    a_t
    \nonumber\\
    &=
        \delta_{tr} a_m
        -
        \delta_{mr} a_t
        +
         a_r^\dagger
         a_m
        a_t.
    \label{eq:local-xxxo-xxox-normal-order}
\end{align}
Substituting Eq.~\eqref{eq:local-xxxo-xxox-normal-order} into
Eq.~\eqref{eq:xxxo-xxox-raw-product} gives
\begin{align}
     A_{xxxo}(I)
     A_{xxox}(J)^\dagger
    ={}&
    \delta_{tr}\,
     a_j^\dagger
     a_k^\dagger
     a_l^\dagger
     a_m
     a_q
     a_p
    \nonumber\\
    &-
    \delta_{mr}\,
     a_j^\dagger
     a_k^\dagger
     a_l^\dagger
     a_t
     a_q
     a_p
    \nonumber\\
    &+
     a_j^\dagger
     a_k^\dagger
     a_l^\dagger
     a_r^\dagger
     a_m
     a_t
     a_q
     a_p.
    \label{eq:xxxo-xxox-full-normal-order}
\end{align}
The first two lines contain six surviving ladder operators and therefore
contribute to $A_6$.  The last line contains eight surviving ladder
operators and therefore contributes to $A_8$.

Let
\begin{equation}
    y_8
    =
    \bigl(
        (\alpha_1,\alpha_2,\alpha_3,\alpha_4);
        (\beta_1,\beta_2,\beta_3,\beta_4)
    \bigr)
\end{equation}
label the canonical degree-eight operator
\begin{equation}
    E_8(y_8)
    =
     a_{\alpha_1}^\dagger
     a_{\alpha_2}^\dagger
     a_{\alpha_3}^\dagger
     a_{\alpha_4}^\dagger
     a_{\beta_4}
     a_{\beta_3}
     a_{\beta_2}
     a_{\beta_1}.
    \label{eq:canonical-E8}
\end{equation}
The degree-eight term in
Eq.~\eqref{eq:xxxo-xxox-full-normal-order} is
\begin{equation}
     a_j^\dagger
     a_k^\dagger
     a_l^\dagger
     a_r^\dagger
     a_m
     a_t
     a_q
     a_p.
\end{equation}
Comparison with Eq.~\eqref{eq:canonical-E8} gives
\begin{equation}
    (\alpha_1,\alpha_2,\alpha_3,\alpha_4)
    =
    (j,k,l,r)
\end{equation}
and
\begin{equation}
    (\beta_1,\beta_2,\beta_3,\beta_4)
    =
    (p,q,t,m).
\end{equation}
Hence the degree-eight permutation map is
\begin{equation}
    P^{(8)}_{xxxo,xxox}
    \bigl(
        (j,k,l,m);
        (p,q,r,t)
    \bigr)
    =
    \bigl(
        (j,k,l,r);
        (p,q,t,m)
    \bigr).
\label{eq:P8-xxxo-xxox}
\end{equation}
The corresponding normal-ordering sign is positive because
$a_r^\dagger$ crosses two annihilation operators,
$a_t$ and $ a_m$.

For a fixed canonical row
\begin{equation}
    y_8
    =
    \bigl(
        (A,B,C,D);
        (P,Q,R,S)
    \bigr),
\end{equation}
the inverse map is obtained by solving
\begin{equation}
    \bigl(
        (A,B,C,D);
        (P,Q,R,S)
    \bigr)
    =
    \bigl(
        (j,k,l,r);
        (p,q,t,m)
    \bigr).
\end{equation}
Therefore,
\begin{equation}
    \left(P^{(8)}_{xxxo,xxox}\right)^{-1}(y_8)
    =
    \bigl(
        (A,B,C,S);
        (P,Q,D,R)
    \bigr).
\label{eq:P8-inverse-xxxo-xxox}
\end{equation}
The contribution of this Gram block to the coefficient of $E_8(y_8)$
is consequently
\begin{equation}
    \left[
        A_{8;xxxo,xxox}G_{xxxo,xxox}
    \right](y_8)
    =
    G_{xxxo,xxox}
    \bigl(
        A,B,C,S;
        P,Q,D,R
    \bigr).
\label{eq:A8-xxxo-xxox}
\end{equation}

Let
\begin{equation}
    y_6
    =
    \bigl(
        (A,B,C);
        (P,Q,R)
    \bigr)
\end{equation}
label the canonical degree-six operator
\begin{equation}
    E_6(y_6)
    =
     a_A^\dagger
     a_B^\dagger
     a_C^\dagger
     a_R
     a_Q
     a_P.
    \label{eq:canonical-E6}
\end{equation}
There are two distinct contraction channels in
Eq.~\eqref{eq:xxxo-xxox-full-normal-order}.

The first degree-six term originating from the contraction $t=r$ is
\begin{equation}
    +
    \delta_{tr}\,
     a_j^\dagger
     a_k^\dagger
     a_l^\dagger
     a_m
     a_q
     a_p.
\end{equation}
To obtain the fixed canonical operator $E_6(y_6)$, the surviving
indices must satisfy
\begin{equation}
    j=A,
    \qquad
    k=B,
    \qquad
    l=C,
    \qquad
    p=P,
    \qquad
    q=Q,
    \qquad
    m=R.
\end{equation}
The Kronecker delta imposes
\begin{equation}
    t=r.
\end{equation}
Introducing a summed dummy index $u$,
\begin{equation}
    t=r=u.
\end{equation}
Thus the index-placement map for this contraction channel is
\begin{equation}
    R^{(6)}_{xxxo,xxox,\kappa_1}(y_6,u)
    =
    \bigl(
        (A,B,C,R);
        (P,Q,u,u)
    \bigr).
\label{eq:iota-first-xxxo-xxox}
\end{equation}
The first contraction channel therefore contributes
\begin{equation}
    \left[
        \mathcal T^{(6)}_{xxxo,xxox,\kappa_1}
        G_{xxxo,xxox}
    \right](y_6)
    =
    \sum_u
    G_{xxxo,xxox}
    \bigl(
        A,B,C,R;
        P,Q,u,u
    \bigr).
\label{eq:first-A6-channel}
\end{equation}

The second degree-six term originating from the sum $m=r$ is
\begin{equation}
    -
    \delta_{mr}\,
     a_j^\dagger
     a_k^\dagger
     a_l^\dagger
     a_t
     a_q
     a_p.
\end{equation}
To obtain $E_6(y_6)$, the surviving indices must satisfy
\begin{equation}
    j=A,
    \qquad
    k=B,
    \qquad
    l=C,
    \qquad
    p=P,
    \qquad
    q=Q,
    \qquad
    t=R.
\end{equation}
The Kronecker delta imposes
\begin{equation}
    m=r=u.
\end{equation}
The corresponding index-placement map is therefore
\begin{equation}
    R^{(6)}_{xxxo,xxox,\kappa_2}(y_6,u)
    =
    \bigl(
        (A,B,C,u);
        (P,Q,u,R)
    \bigr).
\label{eq:iota-second-xxxo-xxox}
\end{equation}
Including the minus sign generated by normal ordering, the second
contraction channel contributes
\begin{equation}
    -
    \left[
        \mathcal T^{(6)}_{xxxo,xxox,\kappa_2}
        G_{xxxo,xxox}
    \right](y_6)
    =
    -
    \sum_u
    G_{xxxo,xxox}
    \bigl(
        A,B,C,u;
        P,Q,u,R
    \bigr).
\label{eq:second-A6-channel}
\end{equation}

Adding the two contraction channels gives the complete contribution of
the $G_{xxxo,xxox}$ Gram block to the coefficient of $E_6(y_6)$:
\begin{equation}
\begin{aligned}
    \left[
        A_{6;xxxo,xxox}G_{xxxo,xxox}
    \right](y_6)
    ={}&
    \sum_u
    G_{xxxo,xxox}
    \bigl(
        A,B,C,R;
        P,Q,u,u
    \bigr)
    \\
    &-
    \sum_u
    G_{xxxo,xxox}
    \bigl(
        A,B,C,u;
        P,Q,u,R
    \bigr).
\end{aligned}
\label{eq:complete-A6-xxxo-xxox}
\end{equation}

The degree-eight branch preserves all eight raw indices.  It is
therefore described by the invertible permutation
$P^{(8)}_{xxxo,xxox}$.

By contrast, each degree-six branch removes one creation operator and
one annihilation operator through a CAR contraction.  The corresponding
two raw indices are identified and summed.  Consequently, there is no
invertible map from the six free indices in $y_6$ to the original eight
raw indices.  Instead, each contraction channel is described by an
index-placement map
\begin{equation}
    R^{(6)}_{\sigma,\tau,\kappa}
    :
    (y_6,u)
    \longmapsto
    (I;J),
\end{equation}
which inserts the six free indices in $y_6$ and the repeated contracted
index $u$ into the eight argument slots of the Gram tensor.

More generally, if
$\mathcal K_{6}(\sigma,\tau)$ denotes the set of normal-ordering
branches of
$A_\sigma(I) A_\tau(J)^\dagger$
that contain six surviving ladder operators, then
\begin{equation}
    \left[A_6G\right](y_6)
    =
    \sum_{\eta}
    \sum_{\sigma,\tau\in\mathcal{S}_\eta}
    \sum_{\kappa\in\mathcal K_{6}(\sigma,\tau)}
    \Gamma_{\sigma,\tau,\kappa}^{(6)}
    \sum_{u_\kappa}
    G_{\sigma,\tau}
    \left(
        R_{\sigma,\tau,\kappa}^{(6)}
        (y_6,u_\kappa)
    \right).
\label{eq:general-symbolic-A6}
\end{equation}
Here
$\Gamma_{\sigma,\tau,\kappa}^{(6)}$ is the CAR coefficient of the
contraction branch, including both its contraction sign and any
additional sign required to place the surviving ladder operators in
the canonical order.  The degree-six truncation condition is
\begin{equation}
    \left[A_6G\right](y_6)=0
    \qquad
    \text{for every canonical degree-six label }y_6.
\end{equation}

\section{Degree-two particle-number multiplier sufficiency at arbitrary SOS level}\label{app:generalized_sufficient_degree2}
The same mechanism works for every $d\geq2$. We first generalize
the CAR identity used to complete a cubic monomial square to the
identity. For a ladder word $W=b_1\cdots b_\ell$, where each $b_j$
is a single creation or annihilation operator and scalar coefficients
are kept separate, we show that $\mathbb{1}-W^\dagger W$ is an SOS
with generator degrees at most $\ell$.

Write $W=bT$, separating the first ladder operator $b=b_1$ from
the remaining word $T=b_2\cdots b_\ell$. Adding and subtracting
$T^\dagger T$, and using
$b^\dagger b+bb^\dagger=\mathbb{1}$, gives
\begin{align}
 \mathbb{1}-W^\dagger W
 &=\mathbb{1}-T^\dagger b^\dagger bT
 \nonumber\\
 &=(\mathbb{1}-T^\dagger T)
   +T^\dagger(\mathbb{1}-b^\dagger b)T
 \nonumber\\
 &=(\mathbb{1}-T^\dagger T)
   +(b^\dagger T)^\dagger(b^\dagger T).
 \label{eq:degree-two-word-complement}
\end{align}
The second term is an explicit square whose generator has length
$\ell$. The first term is the same problem for the shorter word
$T$, of length $\ell-1$. Repeating this operation shortens the
remaining word one ladder at a time, until it is the empty word
$\mathbb{1}$, whose complement is zero. Thus the procedure terminates
after $\ell$ steps without introducing any generator longer than
the original word.

For example, with $W=b_1b_2b_3$, the three steps give
\begin{align}
 \mathbb{1}-W^\dagger W
 ={}&(b_1^\dagger b_2b_3)^\dagger
      (b_1^\dagger b_2b_3)
 \nonumber\\
 &+(b_2^\dagger b_3)^\dagger(b_2^\dagger b_3)
 \nonumber\\
 &+(b_3^\dagger)^\dagger b_3^\dagger.
 \label{eq:degree-two-word-complement-example}
\end{align}
The generators have lengths three, two, and one, respectively.
In general, normal ordering cannot increase these degree bounds,
so
\begin{equation}
 \mathbb{1}-W^\dagger W\in\Sigma_\ell
 \subseteq\Sigma_d,\qquad \ell\leq d.
 \label{eq:degree-two-word-complement-degree}
\end{equation}
This establishes an explicit degree-controlled SOS representation,
rather than only the operator inequality
$W^\dagger W\preceq\mathbb{1}$. The iteration reduces the length
of the remaining word; it does not increase the SOS level. Split each multiplier monomial of degree at most $2d-2$
as $M=U^\dagger V$, with $\deg U\leq d$ and $\deg V\leq d-2$.
For its coefficient $c$, set
$P_\lambda=\lambda^{-1/2}c^*U+\lambda^{1/2}Vr$. Then
\begin{align}
 &cMr+c^*rM^\dagger+\frac{|c|^2}{\lambda}\mathbb{1}
                      +\lambda r^2\nonumber\\
 &\quad=P_\lambda^\dagger P_\lambda
 +\frac{|c|^2}{\lambda}(\mathbb{1}-U^\dagger U)
 +\lambda r(\mathbb{1}-V^\dagger V)r\in\Sigma_d.
\label{eq:degree-two-general-conversion}
\end{align}
Equation~\eqref{eq:degree-two-general-conversion} combines the
same two corrections used in the $(2,3)$ proof, where cubic
generators produce terms of degree at most six. We first complete
the square and isolate a term proportional to $r^2$, which can be
absorbed into a degree-two multiplier. The remaining negative SOS
is then replaced by a positive SOS minus a constant, producing
the small decrease in the certified energy. Here we carry out
these operations monomial by monomial, rather than first grouping
the multiplier coefficients into the polynomials $\phi_i$.

\emph{1. Complete the square and isolate the number penalty.}
For the factorization $M=U^\dagger V$ and generator
$P_\lambda=\lambda^{-1/2}c^*U+\lambda^{1/2}Vr$ introduced above,
direct expansion gives
\begin{align}
 W_N(cM)
 &=cMr+c^*rM^\dagger
 \nonumber\\
 &=P_\lambda^\dagger P_\lambda
   -\lambda rV^\dagger Vr
   -\frac{|c|^2}{\lambda}U^\dagger U.
 \label{eq:degree-two-general-expanded-square}
\end{align}
Both $U$ and $Vr$ have degree at most $d$, so the first term
is an allowed square. The other two terms have negative signs
and must be retained in the operator identity.

To handle the term containing $V$, use the word-complement
construction in Eq.~\eqref{eq:degree-two-word-complement} to write
\begin{equation}
 \mathbb{1}-V^\dagger V
 =\sum_\mu B_\mu^\dagger B_\mu,
 \qquad \deg B_\mu\leq d-2.
\end{equation}
Multiplying on the left and right by the Hermitian operator $r$
gives another explicit SOS:
\begin{equation}
 r(\mathbb{1}-V^\dagger V)r
 =\sum_\mu(B_\mu r)^\dagger(B_\mu r)\in\Sigma_d,
 \label{eq:degree-two-general-number-complement}
\end{equation}
because $\deg(B_\mu r)\leq d$. No commutation of $r$ through
$B_\mu$ is required. Consequently,
\begin{equation}
 -\lambda rV^\dagger Vr
 =\lambda r(\mathbb{1}-V^\dagger V)r-\lambda r^2.
\end{equation}
The first term on the right is now a positive SOS contribution;
the second is the desired particle-number penalty.

This is precisely the role of the particle--hole identity
in Eq.~\eqref{eq:degree-two-car-penalty}. In the $(2,3)$
factorization, $V=a_i$, so $Vr=Z_i$ and
\begin{equation}
 r(\mathbb{1}-V^\dagger V)r
 =r a_i a_i^\dagger r
 =(a_i^\dagger r)^\dagger(a_i^\dagger r)
 =Y_i^\dagger Y_i.
\end{equation}
For a longer word $V$, the single companion square
$Y_i^\dagger Y_i$ is replaced by the sum of squares in
Eq.~\eqref{eq:degree-two-general-number-complement}. The purpose
is unchanged: complete $(Vr)^\dagger(Vr)$ to $r^2$ without
leaving the level-$d$ generator space.

Define
\begin{equation}
 Q_{M,\lambda}
 =P_\lambda^\dagger P_\lambda
  +\lambda r(\mathbb{1}-V^\dagger V)r\in\Sigma_d.
\end{equation}
We have therefore obtained
\begin{equation}
 W_N(cM)
 =Q_{M,\lambda}-\lambda r^2
  -\frac{|c|^2}{\lambda}U^\dagger U.
 \label{eq:degree-two-general-single-residual}
\end{equation}
This has the same structure as
Eq.~\eqref{eq:degree-two-square-completion}: an allowed SOS,
a negative multiple of $r^2$, and a negative SOS residual
whose coefficient decreases as $1/\lambda$.

\emph{2. Absorb the penalty and construct the residual complement.}
Fix a monomial expansion
$f=\sum_{\alpha=1}^{J}c_\alpha M_\alpha$, with
$M_\alpha=U_\alpha^\dagger V_\alpha$ and the degree bounds above.
Applying Eq.~\eqref{eq:degree-two-general-single-residual} to
each term, using the same $\lambda>0$, gives
\begin{align}
 W_N(f)
 &=Q_\lambda^{(d)}-J\lambda r^2-\frac{K_f}{\lambda},
 \label{eq:degree-two-general-summed-residual}\\
 Q_\lambda^{(d)}
 &:=\sum_{\alpha=1}^{J}Q_{M_\alpha,\lambda}\in\Sigma_d,
 \qquad
 K_f:=\sum_{\alpha=1}^{J}|c_\alpha|^2
                         U_\alpha^\dagger U_\alpha.
 \nonumber
\end{align}
Here $J$ counts monomial terms, not modes or particles.
For a starting certificate $H-E\mathbb{1}=S+W_N(f)$,
the penalty is absorbed by the degree-at-most-two multiplier
\begin{equation}
 \widetilde f=-\frac{J\lambda}{2}r
 =\frac{J\lambda\eta}{2}\mathbb{1}
  -\frac{J\lambda}{2} N,
 \qquad
 W_N(\widetilde f)=-J\lambda r^2.
 \label{eq:degree-two-general-multiplier-update}
\end{equation}
Unlike the earlier proof, which converted only $f_{(4)}$ and
retained $f_{\leq2}$, here we convert the entire $f$.
The resulting multiplier therefore consists solely of a
multiple of $r$.

The residual $-K_f/\lambda$ remains and cannot be discarded.
It plays the same role as $-K/\lambda$ in the ${}_{\text{wn}}\Sigma_{3,\leq 2}$ proof.
There we constructed $C\mathbb{1}-K\in\Sigma_3$ after grouping
coefficients into $\phi_i$. Here the monomial-by-monomial
construction gives the corresponding complement directly:
\begin{align}
 C_f&:=\sum_{\alpha=1}^{J}|c_\alpha|^2,
 \nonumber\\
 D_f&:=C_f\mathbb{1}-K_f
 \nonumber\\
 &=\sum_{\alpha=1}^{J}|c_\alpha|^2
       (\mathbb{1}-U_\alpha^\dagger U_\alpha)
 \in\Sigma_d.
 \label{eq:degree-two-general-residual-complement}
\end{align}
Each summand is an explicit SOS by the word-complement
construction, since $\deg U_\alpha\leq d$. Thus this is the
analogue of Eq.~\eqref{eq:degree-two-residual-complement},
not merely an operator-norm bound. The constants obtained by
the two constructions need not agree. What matters is that
$C_f$ is finite and fixed by the chosen starting multiplier
expansion, independently of $\lambda$.

\emph{3. Assemble the certificate and take the supremum.}
Substituting $-K_f/\lambda=D_f/\lambda-(C_f/\lambda)\mathbb{1}$
into Eq.~\eqref{eq:degree-two-general-summed-residual} yields
\begin{equation}
 H-\left(E-\frac{C_f}{\lambda}\right)\mathbb{1}
 =\widetilde S+W_N(\widetilde f),
 \label{eq:degree-two-general-converted-certificate}
\end{equation}
where
\begin{equation}
 \widetilde S
 =S+Q_\lambda^{(d)}+\frac{D_f}{\lambda}\in\Sigma_d.
\end{equation}
All identities are exact at every finite $\lambda>0$.
The only loss in the certified energy is the constant
$C_f/\lambda$; no operator residual is neglected.

Using the definition of $E_{d,m}^*$ in Eq.~\eqref{eq:supremal_energy_def} for the level-$d$ supremal bound with
multiplier cutoff $m$, the constructed certificate gives
\begin{equation}
 E-\frac{C_f}{\lambda}
 \leq E_{d,2}^*
 \leq E_{d,2d-2}^*.
\end{equation}
Keep the original certificate and its constant $C_f$ fixed.
Since the construction works for every $\lambda>0$, taking
$\lambda\to\infty$ gives $E_{d,2}^*\geq E$.
Taking the supremum over all feasible starting values $E$
then establishes equality with the full-multiplier problem.
Moreover, Eq.~\eqref{eq:degree-two-general-multiplier-update}
shows that the one-parameter family
$\widetilde f=-\tau r/2$, $\tau\geq0$, already suffices:
\begin{equation}
 \begin{aligned}
 E_{d,2d-2}^*=E_{d,2}^*
 &=\sup_{\substack{E\in\mathbb R\\\tau\geq0}}
 \left\{E:\ H-E\mathbb{1}+\tau r^2\in\Sigma_d\right\}.
 \end{aligned}
 \label{eq:degree-two-general-equality}
\end{equation}
This concerns equality of suprema, not convergence of the
constructed Gram or multiplier coefficients.

The construction remains compatible with particle-number
Gram blocks: for a particle-conserving monomial
$M=U^\dagger V$, the generators $U$ and $Vr$ have matching
particle-number changes. The complement squares also have
definite charges. The full generator space is nevertheless
needed to guarantee that all these squares are available.
The conversion is independent of the Hamiltonian coefficients
and also applies when $\deg H\leq2d$. For a two-body Hamiltonian,
Eq.~\eqref{eq:degree-two-general-converted-certificate} implies
that the total degree-greater-than-four component of
$\widetilde S$ vanishes exactly, just as the sextic component
did in the $(2,3)$ proof. Finally, the argument keeps $d$
fixed: higher multiplier degrees are redundant in supremal
value, but increasing the SOS level can still strengthen
the bound.

\section{Molecular symmetries}\label{app:gram_symmetries}
The non-relativistic electronic structure Hamiltonian commutes with $S_z$ and $S^{2}$ operators and thus it is common to construct symmetry adapted theories that automatically obey these symmetries. 
Symmetry enters the construction in two distinct ways. Requiring the witness to commute with the conserved quantities allows the Gram matrix to be organized into blocks labeled by the corresponding changes in quantum numbers. Restricting positivity to prescribed eigenvalues is accomplished by including the associated equality-constraint multipliers. For particle number and spin projection, the Gram blocks are labeled by \((\Delta N,2\Delta S_z)\), while the equality constraints are \(N-\eta\mathbb 1\) and \(S_z-s\mathbb 1\).
Let $2\Delta S_z = \Delta N_{\alpha} - \Delta N_{\beta}$. $N_{\alpha/\beta}$ counts the number of spin-up and spin-down SOS generators, respectively. More specifically, for a spin-labeled generator $A_{\sigma}$
\begin{align}
\Delta N_{\alpha}\left(A_{\sigma}\left(I\right)\right) = n_{x\alpha}\left(A_{\sigma}(I) \right) - n_{o\alpha} \left(A_{\sigma}(I) \right)  \\
\Delta N_{\beta}\left(A_{\sigma}\left(I\right)\right) = n_{x\beta} \left(A_{\sigma}(I) \right)  - n_{o\beta}\left(A_{\sigma}(I) \right) 
\end{align}
where $x/o$ labels the existence of the Hermitian conjugate like before. 
\begin{table}[H]
\caption{Quadratic-generator Gram blocks resolved by particle-number
and spin-projection changes, retaining the unreduced operator orderings
of Ref.~\cite{rubin2026near}. Charges are assigned to the displayed
monomials in the $R^{\dagger}R$ convention. Each row defines one
positive-semidefinite Gram block; the last column gives its array of
subblocks associated with the displayed monomial families, not its
full matrix dimensions. Spatial orbital indices $i,j$ are unrestricted
except that $i<j$ for the same-spin pair generators. Block labels and
equation numbers refer to that work.}
\label{tab:sz_gram_blocks}
\begin{ruledtabular}
\begin{tabular}{lccc}
Generator monomials
& $(\Delta N,2\Delta S_z)$
& Gram block
& Subblock array \\
\hline
\begin{tabular}[t]{@{}l@{}}
$a^{\dagger}_{j\alpha}a_{i\alpha},\quad
 a^{\dagger}_{j\beta}a_{i\beta},$\\
$a_{j\alpha}a^{\dagger}_{i\alpha},\quad
 a_{j\beta}a^{\dagger}_{i\beta}$
\end{tabular}
& $(0,0)$
& $G$, Eq.~(36)
& $4\times4$ \\[4pt]
$a^{\dagger}_{j\beta}a_{i\alpha},\quad
 a_{j\alpha}a^{\dagger}_{i\beta}$
& $(0,-2)$
& $G_{\alpha\beta}$, Eq.~(37)
& $2\times2$ \\[4pt]
$a^{\dagger}_{j\alpha}a_{i\beta},\quad
 a_{j\beta}a^{\dagger}_{i\alpha}$
& $(0,+2)$
& $G_{\beta\alpha}$, Eq.~(38)
& $2\times2$ \\[4pt]
$a_{j\alpha}a_{i\alpha}$
& $(-2,-2)$
& $D_{\alpha\alpha}$
& $1\times1$ \\[4pt]
$a_{j\beta}a_{i\beta}$
& $(-2,+2)$
& $D_{\beta\beta}$
& $1\times1$ \\[4pt]
$a_{j\beta}a_{i\alpha},\quad a_{i\alpha}a_{j\beta}$
& $(-2,0)$
& $D_{\alpha\beta}$, Eq.~(39)
& $2\times2$ \\[4pt]
$a^{\dagger}_{j\alpha}a^{\dagger}_{i\alpha}$
& $(+2,+2)$
& $Q_{\alpha\alpha}$
& $1\times1$ \\[4pt]
$a^{\dagger}_{j\beta}a^{\dagger}_{i\beta}$
& $(+2,-2)$
& $Q_{\beta\beta}$
& $1\times1$ \\[4pt]
$a^{\dagger}_{j\beta}a^{\dagger}_{i\alpha},\quad
 a^{\dagger}_{i\alpha}a^{\dagger}_{j\beta}$
& $(+2,0)$
& $Q_{\alpha\beta}$, Eq.~(40)
& $2\times2$ \\
\end{tabular}
\end{ruledtabular}
\end{table}

\end{document}